\documentstyle[preprint,eqsecnum,aps]{revtex}
\begin{document}
\draft
\title{SECOND--POST-NEWTONIAN\\
GENERATION OF GRAVITATIONAL RADIATION}
\author{Luc BLANCHET}
\address{D\'epartement d'Astrophysique Relativiste et de Cosmologie,\\
CNRS (UPR 176), Observatoire de Paris,\\
92195 Meudon Cedex, France}
\date{\today}
\maketitle
\begin{abstract}
This paper derives the expressions of the multipole moments of an
isolated gravitating source with an accuracy corresponding to the second
post-Newtonian (2-PN) approximation of general relativity.  The moments
are obtained by a procedure of matching of the external gravitational
field of the source to the inner field, and are found to be given by
integrals extending over the stress-energy distribution of the matter
fields and the gravitational field.  Although this is not manifest on
their expressions, the moments have a compact support limited to the
material source only (they are thus perfectly well-defined
mathematically).  From the multipole moments we deduce the
expressions of the asymptotic gravitational waveform and associated
energy generated by the source at the 2-PN approximation.  This work,
together with a forthcoming work devoted to the application to coalescing
compact binaries, will be used in the future
observations of gravitational radiation by laser interferometric
detectors.

\end{abstract}
\newpage
\section*{Content}

\noindent 1. INTRODUCTION\\
A. Motivation and relation to other works\\
B. Notation for Einstein's equations\\

\noindent 2. THE INTERNAL AND EXTERNAL GRAVITATIONAL FIELDS\\
A. Solution of Einstein's equations in the internal near zone\\
B. Solution of Einstein's equations in the external near zone\\

\noindent 3. MATCHING OF THE INTERNAL AND EXTERNAL FIELDS\\
A. Coordinate transformation between the internal and external fields\\
B. Matching of the compact-supported potentials $V$ and $V_i$ \\
C. Matching of the non-compact-supported potential $W_{ij}$\\
D. Matching equation at the post-Newtonian order $\varepsilon^6$\\

\noindent 4. GENERATION OF GRAVITATIONAL WAVES\\
A. Relations between the canonical moments and the source moments\\
B. Expressions of the mass-type and current-type source moments\\
C. The asymptotic waveform at the 2-PN approximation\\

\noindent APPENDICES\\
A. PN expansion of part of the external field\\
B. The conserved 2-PN total mass\\
C. Computation of three constants\\

\section{Introduction}
\label{sec:1}

\subsection{Motivation and relation to other works}
\label{ssec:1.1}

The problem of the generation of gravitational radiation by a slowly
moving isolated system is presently solved with an accuracy
corresponding to the 1.5-post-Newtonian (in short 1.5-PN) approximation
of general relativity \cite{BD89,DI91,BD92}.  This approximation takes
into account all contributions in the radiation field at large distances
from the system up to the relative order $\varepsilon^{3}$, where
$\varepsilon \sim c^{-1}$ is a usual post-Newtonian parameter.  The
objective of this paper is to go one step further in the resolution of
this problem by deriving the expression of the gravitational radiation
field with an accuracy corresponding to the 2-PN approximation, taking
into account all contributions up to the order $\varepsilon^{4}$.  In a
forthcoming paper \cite{BDIWW94} we shall consider specifically the case
of the radiation generated by a coalescing binary system made of compact
stars (neutron stars or black holes).

It is now well-known that post-Newtonian effects in the radiation generated
by a coalescing compact binary system should be detectable by future
gravitational-wave interferometers like LIGO and VIRGO
\cite{KSc87,K89,LW90,3mn,FCh93,CF94}.  It has even been realized in
recent years that an extremely accurate theoretical signal is required
in order to achieve the full potential precision on the measurement of
the binary's parameters \cite{3mn}.  This remarkable state of affairs is
made possible by the fact that the signal will spend hundreds to
thousands of cycles in the detector bandwidth, and that as a result the
instantaneous {\it phase} of the signal will be amenable to a very
precise determination.  The main influence of the higher-order
post-Newtonian corrections is thus located in the phase of the signal,
which is itself mainly determined by the rate of decay of the binary's
orbit resulting from the gravitational radiation reaction.  Note that
this potential high precision on the extraction of the binary's
parameters is interesting not only for doing astrophysical measurements
of masses, spins, etc. of stars, but also for testing some aspects of the
nonlinear structure of general relativity \cite{BSat94b}.

Analytical and numerical computations of the radiation generated by a
small mass in circular orbit around a large one have indicated that
post-Newtonian corrections beyond the 3-PN approximation may be needed
\cite{P93,CFPS93,TNaka94,TSasa94}.  The 2-PN approximation worked out
in this paper and its sequel \cite{BDIWW94} thus does not yet reach the
ideal precision required by the observations of coalescing binaries, but
still represents an appreciable improvement of the existing situation.

The investigation of this paper will be based on a
``multipolar-post-Minkowskian" formalism for dealing with the
gravitational field in the external vacuum region of the source.  Such a
formalism combines a multipole moment expansion, valid in the exterior
of the source, with a post-Minkowskian expansion, i.e.  a nonlinearity
expansion or expansion in powers of Newton's constant $G$, valid
wherever the field is weak.  This double expansion formalism is
originally due to Bonnor and collaborators \cite{Bo59,BoR61,BoR66,HR69},
and has been later refined and clarified by Thorne \cite{Th80}.  In
recent years, the formalism has been implemented in an explicit and
constructive way \cite{BD86,B87} (with the help of some mathematical
tools like analytic continuation), and applications have been made to the
problems of gravitational radiation reaction \cite{BD88,B93} and wave
generation \cite{BD89,DI91,BD92}.  The field being determined only in
the {\it exterior} of the source, the multipolar-post-Minkowskian
formalism must be supplemented by a method of matching to the field {\it
inside} the source.  We shall use a variant of the well-known method of
matching of asymptotic expansions (see e.g.  \cite{Lagerst}) which has
already, on several occasions, found its way in general relativity
\cite{Th69,Bu71,K80,BD84}.

Inspection of the solution of the wave generation problem at the 1.5-PN
approximation \cite{BD89,DI91,BD92} readily shows what is needed for
extension at the 2-PN approximation. Indeed, the works \cite{BD89} and
\cite{DI91} have obtained respectively the expressions of the mass-type
and current-type multipole moments of the source with relative precision
$\varepsilon^{2}$, the next-order correction being of order
$O(\varepsilon^{4})$.  The inclusion done in \cite{BD89} of the terms
$\varepsilon^{2}$ in the mass multipole moments permits to solve the
wave generation problem at the 1-PN approximation (in fact, only the
terms $\varepsilon^{2}$ in the quadrupole mass moment are needed).  This
has set on a solid (and well-defined) footing previous works by Epstein
and Wagoner \cite{EW75}, and Thorne \cite{Th80} who obtained formally
correct but divergent expressions of the moments at the 1-PN
approximation (see \cite{BD89} for discussion).  In the work \cite{BD92}
the nonlinear effects in the radiation field were added and shown to
arise at the level $\varepsilon^{3}$, essentially due to the ``tails" of
gravitational waves.  Higher-order nonlinear effects were shown to be of
order $O(\varepsilon^{5})$ at least.  Now recall that the contribution
of a moment with multipolarity $l$ scales like $\varepsilon^{l}$ in the
radiation field, and that the current moments always carry an additional
factor $\varepsilon$ as compared with the corresponding mass moments.
Therefore we conclude that what is needed for solving the 2-PN wave
generation problem is only to find the extension of the expression of the
{\it mass quadrupole} moment of the source with relative precision
$\varepsilon^{4}$.  Note that the computation of the moments of the
source can be done in the {\it near zone} of the source.

This paper will thus mainly focus on the {\it matching} between the external
and internal gravitational fields in the near zone of the source (both
fields are expressed in the form of a post-Newtonian expansion when
$\varepsilon \to 0$) with a precision consistent with the inclusion of
the terms $\varepsilon^{4}$ in the mass multipole moments.  (Although
the quadrupole mass moment is sufficient for our purpose, we shall compute
the terms $\varepsilon^{4}$ in all mass moments of arbitrary multipolarity
$l$.) The matching performed in \cite{BD89} was based on a particularly
simple closed form of the internal gravitational field of the source
including 1-PN corrections.  On the other hand, the matching
performed in \cite{DI91} made use of some specific distributional
kernels to deal with the quadratic nonlinearities of Einstein's
equations.  None of these methods can be applied to our problem, which
necessitates the inclusion of the full 2-PN corrections in the field,
depending not only on quadratic but also on {\it cubic} nonlinearities
of Einstein's equations.  In this paper we shall employ a matching
procedure which is far more general than the ones followed in
\cite{BD89,DI91}.  In particular, we shall show how one can deal with the
cubic nonlinearities of Einstein's equations without any use of
distributional kernels.

The end result obtained in this paper expresses the multipole moments of
the source as integrals extending over the distribution of stress-energy
of the matter fields in the source {\it and} of the gravitational field.
This result is similar to the one we would obtain by using as the source
of the radiation field the total stress-energy (pseudo) tensor of the
matter and gravitational fields, and by considering formally that this
tensor has a spatially compact support limited to the material source.
It is well-known that by proceeding in this formal way (i.e., {\it \`a la}
Epstein-Wagoner-Thorne), we obtain integral expressions of the multipole
moments which are {\it divergent}, because of the non-compact-supported
distribution of the gravitational field.  (Indeed, the
integral expression of a moment with multipolarity $l$ involves a large
power $\sim |{\bf x}|^l$ of the radial distance to the source, which
blows up at infinity.) We prove in this paper that the correct
expressions of the multipole moments must involve also an analytic
continuation factor $|{\bf x}|^B$, where $B$ is a complex number, and
a ``finite part at $B=0$" prescription to deal with the {\it a priori}
bad behavior of the integrals at spatial infinity.  In such a way the
expressions of the moments are perfectly well-defined mathematically.
When a pole $\sim 1/B$ occurs in an integral due to the bad behavior of
the integral at spatial infinity, the ``finite part at $B=0$" introduces
an additional contribution to the moment which must absolutely
be taken into account.  (However, we shall see that no poles occur at
the level investigated in this paper.)

Let us stress that the expressions presented here of the multipole
moments are not {\it manifestly} of compact-supported
form (contrarily to say the lowest-order expressions at the 1-PN level
\cite{BD89}), but are numerically equal, thanks to the properties of the
analytic continuation, to some compact-supported expressions which could
be constructed at the price of introducing more complicated potentials
besides the usual Poisson integrals of the mass and current densities in
the source. Such a construction is however unnecessary (and is somewhat
awkward) in practical computations of the moments for specific sources.

It is likely that the expressions of the moments expressed in this way as
integrals extending over the total stress-energy distribution of the
matter and gravitational fields and regularized by means of analytic
continuation, will admit a generalization to higher nonlinearities in
the field, and higher post-Newtonian approximations.  Such a
generalization, which will be the subject of future work, should permit
the resolution, at least in principle, of the problem of the generation
of gravitational radiation by a general isolated system up to the high
level of approximation required by the observations of coalescing
compact binaries.

This paper is organized as follows.  In section~\ref{sec:2} we
compute the gravitational field both in the near zone of
the source (where a direct post-Newtonian expansion of the field
equations is performed), and in the external near zone of the source
(where we use the multipolar-post-Minkowskian solution of the vacuum
equations).  In section~\ref{sec:3} we impose that the two
fields are isometric in their common domain of validity,
namely the external near zone of the source. This yields a matching
equation valid up to an appropriate post-Newtonian order, and which is used
in section~\ref{sec:4} to obtain the expressions of both the mass and
current multipole moments of the source.  The formulas for the
asymptotic waveform and the associated energy generated by the source at
the 2-PN approximation are also given in section~\ref{sec:4}.  The paper
ends with three appendices, and we start with our notation for
Einstein's equations.

\subsection{Notation for Einstein's equations}
\label{sec:1.2}

Throughout most of this paper, we shall use Einstein's equations in
harmonic coordinates. The field deviation from Minkowski's metric is
denoted by $h^{\alpha\beta}=\sqrt{-g} g^{\alpha\beta} - \eta^{\alpha\beta}$,
where greek indices $\alpha, \beta, \mu, \nu...$ range from 0 to 3,
$g_{\alpha\beta}$ is the usual covariant metric, $g^{\alpha\beta}$ is
the inverse of $g_{\alpha\beta}$, $g$ is the determinant of
$g_{\alpha\beta}$, and $\eta^{\alpha\beta} = {\em diag} (-1,1,1,1)$ is
the Minkowski metric.  The condition of harmonic coordinates reads as
\begin{equation}
 \partial_\beta h^{\alpha\beta} = 0\ , \label{eq:1.1}
\end{equation}
where $\partial_\beta = \partial/\partial x^\beta$ is the usual
derivation with respect to the harmonic coordinates. Einstein's
equations reduced by the condition (\ref{eq:1.1}) are written in the
form
\begin{equation}
\Box h^{\alpha\beta} = {16\pi G\over c^4} \lambda T^{\alpha\beta} +
\Lambda^{\alpha\beta} (h)\ . \label{eq:1.2}
\end{equation}
We denote by $\Box =\eta^{\alpha\beta} \partial_\alpha \partial_\beta$
the {\it flat} d'Alembertian operator, by $T^{\alpha\beta}$ the
stress-energy tensor of the non-gravitational fields ($T^{\alpha
\beta}$ has the dimension of an energy density), and by $\lambda$
the absolute value of $g$, i.e. $\lambda = |g| =-g$. In terms of a
series expansion in the field variable $h^{\alpha\beta}$, we have
\begin{equation}
\lambda = 1 + h + {1\over 2} (h^2 - h_{\mu\nu} h^{\mu\nu}) + O(h^3)\ ,
 \label{eq:1.3}
\end{equation}
where $h_{\mu\nu} = \eta_{\mu\alpha} \eta_{\nu\beta} h^{\alpha\beta}$
and $h =\eta_{\alpha\beta} h^{\alpha\beta}$. The second term in
(\ref{eq:1.2}) is an effective gravitational nonlinear source including
all the nonlinearities (quadratic, cubic, \dots) of Einstein's equations.
We denote
\begin{equation}
 \Lambda^{\alpha\beta} (h) = N^{\alpha\beta} (h,h) + M^{\alpha\beta}
(h,h,h) + O(h^4)  \label{eq:1.4}
\end{equation}
where the quadratically nonlinear term is given by
\begin{eqnarray}
 N^{\alpha\beta} (h,h) =&&- h^{\mu\nu} \partial_\mu \partial_\nu
 h^{\alpha\beta} + {1\over 2} \partial^\alpha h_{\mu\nu} \partial^\beta
 h^{\mu\nu} - {1\over 4} \partial^\alpha h \partial^\beta h \nonumber\\
&&-2 \partial^{(\alpha} h_{\mu\nu} \partial^\mu h^{\beta)\nu}
  +\partial_\nu h^{\alpha\mu} (\partial^\nu h^\beta_\mu + \partial_\mu
  h^{\beta\nu}) \nonumber \\
&&+ \eta^{\alpha\beta} \left[ -{1\over 4}\partial_\rho h_{\mu\nu}
  \partial^\rho h^{\mu\nu} +{1\over 8}\partial_\mu h \partial^\mu h
  +{1\over 2}\partial_\mu h_{\nu\rho} \partial^\nu h^{\mu\rho}\right]\ ,
\label{eq:1.5}
\end{eqnarray}
and where the cubically nonlinear term is given by
\begin{eqnarray}
M^{\alpha\beta} (h,h,h) =&& - h^{\mu\nu} (\partial^\alpha h_{\mu\rho}
 \partial^\beta h^\rho_\nu + \partial_\rho h^\alpha_\mu \partial^\rho
h^\beta_\nu -\partial_\mu h^\alpha_\rho \partial_\nu h^{\beta\rho})\nonumber\\
&& + h^{\alpha\beta} \left[ -{1\over 4}\partial_\rho h_{\mu\nu}
  \partial^\rho h^{\mu\nu} +{1\over 8}\partial_\mu h \partial^\mu h
+{1\over 2}\partial_\mu h_{\nu\rho} \partial^\nu h^{\mu\rho}\right]\nonumber\\
&& +{1\over 2}h^{\mu\nu} \partial^{(\alpha} h_{\mu\nu} \partial^{\beta)}h
  +2 h^{\mu\nu}\partial_\rho h^{(\alpha}_\mu \partial^{\beta)} h^\rho_\nu
 \nonumber \\
&& + h^{\mu(\alpha} \left( \partial^{\beta)} h_{\nu\rho}
 \partial_\mu h^{\nu\rho} -2\partial_\nu h^{\beta)}_\rho \partial_\mu
 h^{\nu\rho} -{1\over 2}\partial^{\beta)} h \partial_\mu h \right)\nonumber\\
&& +\eta^{\alpha\beta}\left[ {1\over 8} h^{\mu\nu}\partial_\mu h\partial_\nu h
 - {1\over 4} h^{\mu\nu}\partial_\rho h_{\mu\nu} \partial^\rho h
 - {1\over 4} h^{\rho\sigma}\partial_\rho h_{\mu\nu} \partial_\sigma h^{\mu\nu}
\right.  \nonumber\\
&& \qquad\left. - {1\over 2} h^{\rho\sigma} \partial_\mu h_{\rho_\nu}
\partial^\nu h^\mu_\sigma + {1\over 2} h^{\rho\sigma} \partial_\mu
h^\nu_\rho \partial^\mu h_{\sigma\nu} \right] \ .
\label{eq:1.6}
\end{eqnarray}
In (\ref{eq:1.5}) and (\ref{eq:1.6}), we raise and lower all indices with
the Minkowski metric,  and we denote $t_{(\alpha\beta)} ={1\over 2}
(t_{\alpha\beta} +t_{\beta\alpha})$. By taking the divergence of
(\ref{eq:1.5}) and (\ref{eq:1.6}), we obtain the relations
\begin{eqnarray}
\partial_\beta N^{\alpha\beta} &=& \left( -\partial_\mu h^\alpha_\nu +
{1\over 2} \partial^\alpha h_{\mu\nu} - {1\over 4} \partial^\alpha
           h\, \eta_{\mu\nu} \right) \Box h^{\mu\nu}\ , \label{eq:1.7}\\
\partial_\beta M^{\alpha\beta} &=& \left( \partial_\mu h^\alpha_\nu -
{1\over 2} \partial^\alpha h_{\mu\nu} + {1\over 4} \partial^\alpha
  h\, \eta_{\mu\nu} \right) N^{\mu\nu} \nonumber \\
 &&+ \left[ {1\over 2} h^{\alpha\rho} \partial_\rho h_{\mu\nu} -
 h_{\mu\rho}\partial^\alpha h^\rho_\nu +h_{\mu\rho}\partial_\nu h^{\alpha\rho}
 \right.\nonumber \\ && \qquad\left.
 +{1\over 4} h_{\mu\nu} \partial^\alpha h + {1\over 4} h_{\rho\sigma}
\partial^\alpha h^{\rho\sigma} \eta_{\mu\nu} - {1\over 4} h^{\alpha\rho}
\partial_\rho h\, \eta_{\mu\nu} \right] \Box h^{\mu\nu}\ , \label{eq:1.8}
\end{eqnarray}
which are consistent, by Bianchi's identities, with the conservation (in
the covariant sense) of the stress-energy tensor of the matter fields.

\section{The internal and external gravitational fields}
\label{sec:2}

\subsection{Solution of Einstein's equations in the internal near zone}
\label{ssec:2.1}

Let us define the internal near zone of the source to be a domain $D_i=
\{({\bf x},t)/|{\bf x}|< r_i\}$ whose radius $r_i$ is adjusted so that
(i) $r_i>a$, where $a$ is the radius of a sphere which totally encloses
the source, and
(ii) $r_i \ll \lambdabar$, where $\lambdabar \sim ac/v$ is the (reduced)
wavelength of the emitted gravitational radiation, and $v$ is a typical
internal velocity in the source.

Defining $D_i$ in such a way assumes in particular that $a\ll \lambdabar$,
or equivalently $\varepsilon \ll 1$ where $\varepsilon \sim v/c$ is a
small ``post-Newtonian" parameter appropriate to the description of
slowly moving sources. (We shall also assume that the source is
self-gravitating so that $GM/(ac^2) \sim \varepsilon^2$ where $M$
is the total mass of the source, and that the internal stresses are such
that $T^{ij}/T^{00}\sim \varepsilon^2$.) In $D_i$, we can solve Einstein's
equations (\ref{eq:1.1})-(\ref{eq:1.2}) by formally taking the
post-Newtonian limit $\varepsilon\to 0$. The following notation will be
used for terms of small order in the post-Newtonian parameter
$\varepsilon$. By $A=O(p)$ we mean that $A$ is of order
$O(\varepsilon^p)$; by $B^\alpha = O(p,q)$ we mean that the zero
component of the vector $B^\alpha$ is $B^0 =O(p)$, and that the spatial
components of $B^\alpha$ are $B^i =O(q)$; and similarly by
$C^{\alpha\beta} =O(p,q,r)$ we mean that $C^{00} = O(p)$, $C^{0i} =O(q)$
and $C^{ij} =O(r)$. (The latin indices $i,j$,\dots range from 1 to 3.)

{}From the {\it contravariant} components of the stress energy tensor
$T^{\alpha\beta}$ of the matter fields, we define a mass density $\sigma$,
a current density $\sigma_i$ and a stress density $\sigma_{ij}$ by the
formulas
\begin{mathletters}
\label{eq:2.1}
\begin{eqnarray}
 \sigma &=& {T^{00} + T^{ii}\over c^2} \ , \label{eq:2.1a}\\
 \sigma_i &=& {T^{0i}\over c} \ , \label{eq:2.1b}\\
 \sigma_{ij} &=& T^{ij} \ , \label{eq:2.1c}
\end{eqnarray}
\end{mathletters}
where in (\ref{eq:2.1a}) $T^{ii} =\Sigma \delta_{ij} T^{ij}$ denotes the
spatial trace of $T^{\alpha\beta}$. In all this paper, we shall assume
that the matter densities $\sigma$, $\sigma_i$ and $\sigma_{ij}$ are of
order $\varepsilon^0$ when $\varepsilon \to 0$, i.e.
\begin{equation}
 \sigma, \ \sigma_i,\ \sigma_{ij} = O(0)\ . \label{eq:2.2}
\end{equation}
We introduce next some {\it retarded} potentials generated by the
densities $\sigma$, $\sigma_i$ and $\sigma_{ij}$. First, $V$ and $V_i$
are the usual retarded scalar and vector potentials of the mass and
current densities $\sigma$ and $\sigma_i$, i.e.
\begin{mathletters}
\label{eq:2.3}
\begin{eqnarray}
 V({\bf x},t) &=& G \int {d^3{\bf x}'\over |{\bf x}-{\bf x}'|}\ \sigma
  ({\bf x}', t - {1\over c} |{\bf x}-{\bf x}'|)\ , \label{eq:2.3a}\\
 V_i({\bf x},t) &=& G \int {d^3{\bf x}'\over |{\bf x}-{\bf x}'|}\ \sigma_i
 ( {\bf x}', t - {1\over c} |{\bf x}-{\bf x}'|)\ , \label{eq:2.3b}
\end{eqnarray}
\end{mathletters}
satisfying $\Box V=-4\pi G\sigma$ and $\Box V_i =-4\pi G\sigma_i$. Second,
$W_{ij}$ is a more complicated retarded tensor potential defined by
\FL\begin{equation}
 W_{ij} ({\bf x},t) = G \int {d^3{\bf x}'\over |{\bf x}-{\bf x}'|}
\left[ \sigma_{ij} + {1\over 4\pi G} ( \partial_i V\partial_j V
 - {1\over 2} \delta_{ij} \partial_k V\partial_k V )\right]
 ( {\bf x}', t-{1\over c} |{\bf x}-{\bf x}'|)\ . \label{eq:2.4}
\end{equation}
Note that while the potentials $V$ and $V_i$ have a {\it compact} support
limited to that portion of the past null cone issued from the field
point $({\bf x}',t)$ which intersects the source, the potential $W_{ij}$
has not a compact support. From (\ref{eq:2.3a}), we see that $V({\bf x}',
t-{1\over c} |{\bf x}-{\bf x}'|)$ behaves like $GM/|{\bf x}'|$ when
$|{\bf x}'|\to \infty$, where $M=\int d^3 {\bf y}\sigma ({\bf y}, -\infty)$
$+O(2)$ is the
initial mass (or ADM mass) of the source, and we can check from this that
the potential $W_{ij}$ is given as a convergent integral. We shall often
abbreviate formulas such as (\ref{eq:2.3}-\ref{eq:2.4}) by denoting the
retarded integral of some source $f({\bf x},t)$, having adequate fall-off
properties along past null cones, by
\begin{equation}
 (\Box^{-1}_R f)({\bf x},t) = -{1\over 4\pi} \int {d^3{\bf x}'\over
 |{\bf x}-{\bf x}'|} f ({\bf x}', t-{1\over c} |{\bf x}-{\bf x}'|)\ ,
\label{eq:2.5}
\end{equation}
so that e.g. $V =-4\pi G \Box^{-1}_R \sigma.$

 It is convenient in most of this paper to keep the potentials
(\ref{eq:2.3}-\ref{eq:2.4}) in retarded form, i.e. to {\it not} expand
when $c\to +\infty$ the retardation argument $t- |{\bf x}-{\bf x}'| /c$.
This permits to avoid possible problems of convergence with the expansion
of potentials with non-compact support. However, most of the equations
of this paper will be only valid up to some remainder in the expansion
$\varepsilon \sim c^{-1} \to 0$, and we can replace, when there is no
problem of convergence, the retarded potentials by their post-Newtonian
expanded forms up to the accuracy of the remainder. For instance, from
the Newtonian equations of continuity and of
motion,
\begin{mathletters}
\label{eq:2.6}
\begin{eqnarray}
 \partial_t \sigma + \partial_i \sigma_i &=& O(2)\ , \label{eq:2.6a}\\
 \partial_t \sigma_i+ \partial_j \sigma_{ij} &=& \sigma \partial_i V
    + O(2)\ , \label{eq:2.6b}
\end{eqnarray}
\end{mathletters}
we deduce that the potentials $V$, $V_i$ and $W_{ij}$ satisfy
the conservation laws
\begin{mathletters}
\label{eq:2.7}
\begin{eqnarray}
 \partial_t V + \partial_i V_i &=& O(2)\ , \label{eq:2.7a}\\
 \partial_t V_i + \partial_j W_{ij} &=& O(2)\ . \label{eq:2.7b}
\end{eqnarray}
\end{mathletters}
In (\ref{eq:2.6b}) and (\ref{eq:2.7}) we can replace the potentials $V$,
$V_i$ and $W_{ij}$ by corresponding Poisson-type potentials, e.g.,
$V=U+O(2)$ where $U$ is the Newtonian potential of the mass density
$\sigma$, satisfying $\Delta U = -4\pi G\sigma$.

We now proceed to solve Einstein's equations (\ref{eq:1.1}-\ref{eq:1.2})
with an accuracy corresponding to the post-Newtonian order $O(8,7,8)$,
by which we mean $O(8)$ in the $00$ and $ij$ components of
$h^{\alpha\beta}$, and $O(7)$ in its $0i$ components. The insertion of
the lowest-order results $h^{00} = -4V/c^2 + O(2)$, $h^{0i} =O(3)$ and
$h^{ij} =O(4)$ into the right-hand-side of (\ref{eq:1.2}) with the
explicit expression (\ref{eq:1.5}), yields first the equations to be
solved at order $O(6,5,6)$. We get
\begin{mathletters}
\label{eq:2.8}
\begin{eqnarray}
 \Box h^{00} &=& {16\pi G\over c^4} \left( 1+ {4V\over c^2}\right) T^{00}
 - {14\over c^4} \partial_k V \partial_k V + O(6)\ , \label{eq:2.8a}\\
 \Box h^{0i} &=& {16\pi G\over c^4} T^{0i} + O(5)\ , \label{eq:2.8b}\\
 \Box h^{ij} &=& {16\pi G\over c^4} T^{ij}
 + {4\over c^4} \left\{ \partial_i V \partial_j V - {1\over 2} \delta_{ij}
 \partial_k V \partial_k V \right\} + O(6)\ . \label{eq:2.8c}
\end{eqnarray}
\end{mathletters}
These equations can be straightforwardly solved by means of the
potentials $V$, $V_i$ and $W_{ij}$ we defined in (\ref{eq:2.3}) and
(\ref{eq:2.4}), and by means of the trace $W\equiv W_{ii} \equiv \Sigma
\delta_{ij}W_{ij}$ of the potential $W_{ij}$. The result is
\begin{mathletters}
\label{eq:2.9}
\begin{eqnarray}
h^{00} &=& -{4\over c^2} V +{4\over c^4} (W - 2V^2) +O(6)\ ,\label{eq:2.9a}\\
h^{0i} &=& -{4\over c^3} V_i +O(5)\ ,\label{eq:2.9b}\\
h^{ij} &=& -{4\over c^4} W_{ij} +O(6)\ .\label{eq:2.9c}
\end{eqnarray}
\end{mathletters}
Since the potentials satisfy the conservation laws (\ref{eq:2.7}), we see
that the field (\ref{eq:2.9}) satisfies the approximate harmonic gauge
condition $\partial_\beta h^{\alpha\beta} = O(5,6)$.

The next step consists in iterating Einstein's equations by using the
expression (\ref{eq:2.9}) of the field.  We substitute (\ref{eq:2.9})
into the right-hand-side of (\ref{eq:1.2}) with the help of (\ref{eq:1.3})
and of the explicit expressions (\ref{eq:1.5}) and (\ref{eq:1.6}) of the
quadratic and cubic nonlinearities.  In this way, we find the equations
to the satisfied by the field up to the order $O(8,7,8)$, which are
\begin{equation}
 \Box h^{\alpha\beta} = {16\pi G\over c^4} \overline\lambda (V,W)
  T^{\alpha\beta} + \overline{\Lambda}^{\alpha\beta} (V, V_i, W_{ij}) +
   O(8,7,8)\ . \label{eq:2.10}
\end{equation}
In these equations, we denote by $\overline\lambda$ the post-Newtonian
expansion of $\lambda = |g| =-g$ when {\it truncated} at the order
$O(6)$, i.e. $\lambda =\overline\lambda +O(6)$, and expressed with the
potentials $V$ and $W =W_{ii}$. From (\ref{eq:1.3}) we have
\begin{equation}
 \overline\lambda (V, W) = 1 + {4\over c^2} V - {8\over c^4} (W-V^2)\ .
 \label{eq:2.11}
\end{equation}
(Note that because the matter stress-energy tensor is $T^{\alpha\beta} =
O(-2,-1,0)$, the precision on $\lambda$ as given by (\ref{eq:2.11}) is
necessary only in the 00 component of the equations (\ref{eq:2.10}).)
Similarly, $\overline\Lambda^{\alpha\beta}$ in (\ref{eq:2.10}) is
defined to be the post-Newtonian expansion of the effective nonlinear
source in the right-hand-side of Einstein's equations (\ref{eq:1.2})
when {\it truncated} at order O(8,7,8), i.e. $\Lambda^{\alpha\beta} =
\overline\Lambda^{\alpha\beta} +O(8,7,8)$, and expressed in terms of
the potentials $V$, $V_i$ and $W_{ij}$. From (\ref{eq:1.5}) and
(\ref{eq:1.6}), we arrive at the expressions
\begin{mathletters}
\label{eq:2.12}
\begin{eqnarray}
 \overline\Lambda^{00}(V,V_i, W_{ij}) &=& - {14\over c^4} \partial_k V
 \partial_k V + {16\over c^6} \biggl\{ - V \partial^2_t V - 2V_k \partial_t
 \partial_k V \nonumber \\
 && \qquad - W_{km} \partial_{km}^2 V + {5\over 8} (\partial_t V)^2
 + {1\over 2} \partial_k V_m (\partial_k V_m +3\partial_m V_k) \nonumber\\
 && \qquad + \partial_k V\partial_t V_k + 2\partial_k V\partial_k W
  - {7\over 2} V\partial_k V\partial_k V \biggr\} \ , \label{eq:2.12a} \\
 \overline\Lambda^{0i}(V,V_i, W_{ij}) &=&  {16\over c^5} \left\{ \partial_k V
 (\partial_i V_k - \partial_k V_i) + {3\over 4} \partial_t V \partial_i V
 \right\} \ , \label{eq:2.12b}\\
 \overline\Lambda^{ij}(V,V_i, W_{ij}) &=& {4\over c^4}\left\{ \partial_i V
 \partial_j V -{1\over 2} \delta_{ij} \partial_k V \partial_k V \right\}
  + {16\over c^6} \biggl\{ 2 \partial_{(i} V\partial_t V_{j)}
  - \partial_i V_k \partial_j V_k  \nonumber \\
 && \qquad - \partial_k V_i \partial_k V_j
  + 2 \partial_{(i} V_k \partial_k V_{j)} - {3\over 8} \delta_{ij}
    (\partial_t V)^2 - \delta_{ij} \partial_k V \partial_tV_k \nonumber \\
 && \qquad + {1\over 2} \delta_{ij}\partial_k V_m (\partial_k V_m
   - \partial_m V_k) \biggr\} \ . \label{eq:2.12c}
\end{eqnarray}
\end{mathletters}
By using the equations of motion and of continuity including $\varepsilon^2$
terms (which are given by (\ref{eq:B4}) in the appendix~B
below), one can check that the expressions (\ref{eq:2.11}) and
(\ref{eq:2.12}) imply
\begin{equation}
 \partial_\beta \left[ {16\pi G\over c^4} \overline\lambda T^{\alpha\beta}
  + \overline\Lambda^{\alpha\beta} \right] = O (7,8)\ . \label{eq:2.13}
\end{equation}

Note that the condition (\ref{eq:2.13}) checks all terms in
(\ref{eq:2.12}) except the $c^{-6}$ terms in the $00$ component
$\overline\Lambda^{00}$ of (\ref{eq:2.12a}) and the $c^{-4}$ term in
$\overline\lambda$. We have checked these terms by adding in the $0i$
component $\overline\Lambda^{0i}$ of (\ref{eq:2.12b}) the next order
$c^{-7}$ terms and computing the divergence. Note also that the consideration
of the next order $c^{-7}$ terms in $\overline\Lambda^{0i}$ would allow
the control of the $c^{-4}$ correction terms not only in the mass-type
source moments (which is the aim of this paper), but also in the
current-type source moments. However, we have chosen here not to include
these terms because they are not necessary for solving the 2-PN wave
generation problem, and because they somewhat complicate the discussion
with the need of introducing new potentials besides $V$, $V_i$ and
$W_{ij}$, and the necessity of a better control of the external metric
in \S\ref{ssec:2.2}. These terms will be considered in a future work
where we investigate how the procedure followed in this paper could be
systematically extended to higher post-Newtonian orders.

Finally, a solution of (\ref{eq:2.10}), with the required precision,
can simply be written as
\begin{equation}
 h^{\alpha\beta} = \Box^{-1}_{R} \left[ {16\pi G\over c^4}
\overline\lambda T^{\alpha\beta} + \overline\Lambda^{\alpha\beta} \right]
 + O(8,7,8)\ , \label{eq:2.14}
\end{equation}
where $\Box^{-1}_R$ is the retarded integral operator defined in
(\ref{eq:2.5}). By (\ref{eq:2.13}), this solution satisfies also the
approximate harmonic gauge condition
\begin{equation}
 \partial_\beta h^{\alpha\beta} = O(7,8)\ . \label{eq:2.15}
\end{equation}
Note that we could have added in (\ref{eq:2.14}) some homogeneous
solutions of the wave equation which are {\it regular} in $D_i$ and satisfy
the harmonic gauge condition (\ref{eq:2.15}). It is simpler to use the
solution (\ref{eq:2.14}) as it stands since we shall show that it directly
matches the exterior metric.

\subsection{Solution of Einstein's equations in the external near zone}
\label{ssec:2.2}

Let $D_e =\{({\bf x},t)\,/\,r\equiv |{\bf x}| >r_e\}$ be an external domain
surrounding the source, where $r_e$ is adjusted so that $a<r_e<r_i$, $a$
being the radius of the source and $r_i$ the radius of the inner domain
$D_i$ defined in \S\ref{ssec:2.1}. By our assumption  $GM/(ac^2)
\sim \varepsilon^2$, gravity is weak everywhere and in particular
in $D_e$, so we can solve Einstein's {\it vacuum} equations in $D_e$ by
means of the multipolar and post-Minkowskian approximation method
developed in our previous works \cite{BD86,B87,BD88,B93} on foundations
laid by Bonnor \cite{Bo59} and Thorne \cite{Th80}. More specifically,
we use the construction of the external field which is defined in \S~4.3
of \cite{BD86} and which is referred there to as the ``canonical"
external field. Using a formal infinite post-Minkowskian expansion, or
expansion in powers of Newton's parameter $G$, the ``canonical" external
field $h^{\mu\nu}_{\rm can} = \sqrt{-g_{\rm can}}\ g^{\mu\nu}_{\rm can} -
\eta^{\mu\nu}$ is given as
\begin{equation}
h^{\mu\nu}_{\rm can} = G\,h^{\mu\nu}_{\rm can(1)}+ G^2h^{\mu\nu}_{\rm can(2)}
  +\cdots + G^nh^{\mu\nu}_{{\rm can}(n)} + \cdots , \label{eq:2.16}
\end{equation}
where the coefficients $h^{\mu\nu}_{{\rm can}(n)}$ of an arbitrary $n$th
power of $G$ are {\it algorithmically} constructed from the knowledge
of the previous coefficients $h^{\mu\nu}_{{\rm can}(m)}$ (with $m<n$) by
post-Minkowskian iteration of the vacuum equations. This iteration is
made possible by a systematic use of multipole expansions, valid outside
the source (in $D_e$), for all the coefficients $h^{\mu\nu}_{{\rm
can}(n)}$ in (\ref{eq:2.16}).

The whole iteration in (\ref{eq:2.16}) rests on the first of the
coefficients, namely the ``linearized" field $h^{\mu\nu}_{\rm can(1)}$,
which is chosen to be the most general solution, modulo an arbitrary
linear gauge transformation, of the linear vacuum equations. This solution
reads as
\begin{mathletters}
\label{eq:2.17}
\begin{eqnarray}
 G\,h^{00}_{\rm can(1)} &=&- {4\over c^2}\,V^{\rm ext}\ ,\label{eq:2.17a}\\
 G\,h^{0i}_{\rm can(1)} &=&- {4\over c^3}\,V_i^{\rm ext}\ ,\label{eq:2.17b}\\
 G\,h^{ij}_{\rm can(1)} &=&- {4\over c^4}\,V_{ij}^{\rm ext}\ ,
     \label{eq:2.17c}
\end{eqnarray}
\end{mathletters}
where the external potentials $V^{\rm ext}$, $V^{\rm ext}_i$, $V^{\rm
ext}_{ij}$ (different from the inner potentials $V$, $V_i$, $W_{ij}$)
are given by some explicit infinite multipole expansions of retarded
spherical waves (solutions of the homogeneous wave equation in $D_e$),
namely
\begin{mathletters}
\label{eq:2.18}
\begin{eqnarray}
 V^{\rm ext} &=& G \sum^\infty_{\ell =0} {(-)^\ell\over \ell !}
\partial_L \left[ {1\over r} \,M_L \left( t-{r\over c}\right)\right]\ ,
 \label{eq:2.18a} \\
 V_i^{\rm ext} &=& -G \sum^\infty_{\ell =1} {(-)^\ell\over \ell !}
  \partial_{L-1} \left[ {1\over r} \,M^{(1)}_{iL-1}
  \left( t-{r\over c}\right)\right]\nonumber \\
   && -G \sum^\infty_{\ell =1} {(-)^\ell\over \ell !}{\ell\over \ell +1}
 \varepsilon_{iab} \partial_{aL-1} \left[ {1\over r} \,S_{bL-1}
  \left( t-{r\over c}\right)\right]\ , \label{eq:2.18b}\\
 V_{ij}^{\rm ext} &=& G \sum^\infty_{\ell =2} {(-)^\ell\over \ell !}
  \partial_{L-2} \left[ {1\over r} \,M^{(2)}_{ijL-2}
  \left( t-{r\over c}\right)\right]\nonumber \\
   && +G \sum^\infty_{\ell =2} {(-)^\ell\over \ell !}{2\ell\over \ell +1}
\partial_{aL-2} \left[ {1\over r} \varepsilon_{ab(i}  \,S^{(1)}_{j)bL-2}
  \left( t-{r\over c}\right)\right]\ . \label{eq:2.18c}
\end{eqnarray}
\end{mathletters}
See e.g. (8.12) in Thorne \cite{Th80}. Our notation is as follows
(anticipating also on future needs). Upper case latin letters denote
multi-indices with the corresponding lower case letters being the number
of indices, e.g. $L=i_1 i_2\cdots i_\ell$. Similarly, $L-1=i_1\cdots
i_{\ell -1}$, $L-2=i_1\cdots i_{\ell -2}$ and $aL =ai_1\cdots i_\ell$.
A product of space derivatives $\partial_i =\partial /\partial x^i$ is
denoted by $\partial_L =\partial_{i_1}\partial_{i_2}\cdots\partial_{i_\ell}$.
Similarly, $x^L =x^{i_1} x^{i_2}\cdots x^{i_\ell}$ and $n^L=n^{i_1}
n^{i_2}\cdots n^{i_\ell}$ where $n^i =x^i/|{\bf x}| =x^i/r$. The symmetric
and trace-free (STF) projection is denoted with a hat, e.g.
$\widehat\partial_L$ or $\widehat x^L$, or sometimes by e.g.
$\partial_{\langle L\rangle}$. $M^{(p)}(t)$ denotes the $p$th time
derivative of $M(t)$, and $T_{ij} ={1\over 2} (T_{ij} +T_{ji})$.

The linearized external field (\ref{eq:2.17}-\ref{eq:2.18}) satisfies
the linearized equations $\Box h^{\mu\nu}_{\rm can(1)} =0$ everywhere
except at the spatial origin $r=0$ of the coordinates. The harmonic
gauge condition $\partial_\nu h^{\mu\nu}_{\rm can(1)} =0$ follows from
the (exact) identities
\begin{mathletters}
\label{eq:2.19}
\begin{eqnarray}
\partial_t V^{\rm ext}+\partial_i V_i^{\rm ext} &=& 0\ ,\label{eq:2.19a}\\
\partial_t V_i^{\rm ext}+\partial_j V_{ij}^{\rm ext} &=& 0\ .\label{eq:2.19b}
\end{eqnarray}
\end{mathletters}
Note that the potential $V^{\rm ext}_{ij}$ is trace-free:
\begin{equation}
V^{\rm ext}_{ii} = 0\ . \label{eq:2.20}
\end{equation}
As we see, the potentials (\ref{eq:2.18}) depend on two infinite sets of
functions of time, $M_L(t)$ for $\ell =0,\cdots,\infty$, and $S_L(t)$
for $\ell =1,\cdots,\infty$.  These functions are STF in their $\ell$
indices.  They can be viewed respectively as some ``canonical" mass-type
and current-type multipole moments parametrizing the external canonical
metric.  They are completely arbitrary functions of time except that the
lowest multipole moments $M$ (mass monopole), $M_i$ (mass dipole) and
$S_i$ (current dipole) are constant:  $M^{(1)} = M^{(1)}_i = S^{(1)}_i
=0$.  Note that it was assumed in \cite{BD86} that the moments $M_L(t)$
and $S_L(t)$ are constant before some remote date in the past.  We shall
admit here that one can cover a more general situation where (for
instance) the $\ell$th time-derivatives of $M_L(t)$ and $S_L(t)$ become
asymptotically constant when $t\to -\infty$, corresponding to a
situation of initial scattering in the infinite past.  Furthermore, we
shall relax without justification the assumption made in \cite{BD86}
that the multipole expansions are finite; that is, we assume that we
really have two infinite sets of moments $M_L(t)$ and $S_L(t)$.
These two amendments almost
certainly have no incidence on the results derived in this paper.

 Starting with the linearized metric (\ref{eq:2.17}-\ref{eq:2.18}), one
constructs iteratively the $n$th coefficient $h^{\mu\nu}_{{\rm can}(n)}$
of the series (\ref{eq:2.16}) by the formula
\begin{equation}
h^{\mu\nu}_{{\rm can}(n)} = {\rm FP}_{B=0}\ \Box^{-1}_{R} \left[
 r^B \Lambda^{\mu\nu}_{{\rm can}(n)} \right] + q^{\mu\nu}_{{\rm can}(n)}
 \ . \label{eq:2.21}
\end{equation}
The first term in this formula involves $\Lambda^{\mu\nu}_{{\rm can}(n)}$,
which is defined to be the coefficient of $G^n$ in the expansion of the
effective gravitational source $\Lambda^{\mu\nu} (h)$ in the
right-hand-side of Einstein's equation (\ref{eq:1.2}) and computed with
the ``canonical" field (\ref{eq:2.16}).
That is,
\begin{equation}
 \Lambda^{\mu\nu}(h_{\rm can}) = G^2 \Lambda^{\mu\nu}_{\rm can(2)} + \cdots
 + G^n \Lambda^{\mu\nu}_{{\rm can}(n)} +\cdots . \label{eq:2.22}
\end{equation}
Since $\Lambda^{\mu\nu}$ is at least quadratic in $h$, $\Lambda^{\mu\nu}
_{{\rm can}(n)}$ for any $n$ is a function only of the previous $n-1$
coefficients $h_{{\rm can}(1)},\cdots,h_{{\rm can}(n-1)}$. For instance,
we have (with evident notation)
\begin{mathletters}
\label{eq:2.23}
\begin{eqnarray}
  \Lambda^{\mu\nu}_{\rm can(2)} &=& N^{\mu\nu} (h_{\rm can(1)},
    h_{\rm can(1)})\ , \label{eq:2.23a}\\
  \Lambda^{\mu\nu}_{\rm can(3)} &=& N^{\mu\nu} (h_{\rm can(1)},
    h_{\rm can(2)}) + N^{\mu\nu} (h_{\rm can(2)}, h_{\rm can(1)}) +
   M^{\mu\nu} (h_{\rm can(1)}, h_{\rm can(1)}, h_{\rm can(1)})
    \ , \label{eq:2.23b}
\end{eqnarray}
\end{mathletters}
where $N^{\mu\nu}$ and $M^{\mu\nu}$ are given by
(\ref{eq:1.4})-(\ref{eq:1.6}).  The retarded integral operator
$\Box^{-1}_R$, which is defined in (\ref{eq:2.5}), acts on the source
$\Lambda^{\mu\nu}_{{\rm can}(n)}$ but multiplied by an analytic
continuation factor $r^B$, where $r=|{\bf x}|$ and $B$ is a complex
number.  The introduction of this factor is required because the
linearized metric $h_{\rm can(1)}$ of (\ref{eq:2.17})-(\ref{eq:2.18}),
and all subsequent metrics $h_{{\rm can}(n)}$ and sources $\Lambda_{{\rm
can}(n)}$, are valid only in the exterior of the source (in $D_e$) and
are singular at the spatial origin of the coordinates, $r=0$, located
within the source.  It has been shown in \cite{BD86} that for $B$ a
complex number the retarded integral $f(B) =\Box^{-1}_R (r^B
\Lambda_{{\rm can}(n)})$ defines an analytic function of $B$ all over
the complex plane except in general at integer values of $B$.  Near the
value $B=0$, $f(B)$ admits a Laurent expansion of the
type $f(B)=\Sigma a_p B^p$, where $p\in Z\!\!\!Z$.  The coefficient of
$B^0$ in this expansion, i.e. $a_0$, is what we call the {\it finite part at}
$B=0$ (or ${ \rm FP}_{B=0}$) of the retarded integral $f(B)$.  This is the
first term in (\ref{eq:2.21}); it satisfies $\Box
a_0 = \Lambda_{{\rm can}(n)}$ (and is also singular at the origin).
Thus the introduction of the analytic continuation factor $r^B$ is a
mean (and a convenient one) to obtain a solution of the Einstein
equation (\ref{eq:1.2}) in $D_e$ ---~this is the only thing we need (see
\cite{BD86} for more details about this way of proceeding).  The second
term $q_{{\rm can}(n)}$ in (\ref{eq:2.21}) is a particular retarded
solution of the wave equation, i.e.  $\Box q_{{\rm can}(n)} =0$, whose
divergence is the opposite of the divergence of the first term, and thus
which permits to ensure the satisfaction of the harmonic coordinate
condition $\partial_\nu h^{\mu\nu}_{{\rm can}(n)} =0.$ The precise
definition of the term $q_{{\rm can}(n)}$ is reported in the appendix~A
where we control its order of magnitude in the post-Newtonian expansion.

 The external field (\ref{eq:2.16}), in which we have
(\ref{eq:2.17})-(\ref{eq:2.18}) and (\ref{eq:2.21}), represents the most
general solution of Einstein's equations in $D_e$, and is parametrized
by the arbitrary ``canonical" multipole moments $M_L(t)$ and $S_L(t)$.
Now the point is that one knows (see \cite{B87,BD89,BD92}) how to relate
the observable ``radiative" multipole moments $U_L(t)$ and $V_L(t)$,
parametrizing the field in the distant wave zone of the source (where
the detector is located), to the moments $M_L(t)$ and $S_L(t)$ (see
\S\ref{sec:4} below). Therefore, what is only needed in order to compute the
distant wave field is to consider the near-zone expansion, or expansion
when $c\to+ \infty$ or $\varepsilon \to 0$, of the external field
(\ref{eq:2.16}) up to some suitable order, so as to give by matching to
the inner metric constructed in \S~IIA a suitably accurate physical meaning
to the moments $M_L(t)$ and $S_L(t)$ in terms of the source's parameters.
Equivalently, this means that we must consider the external field
(\ref{eq:2.16}) in that part of the external domain $D_e$ which belongs
to the near zone, i.e. $D_e \cap D_i =\{ ({\bf x},t)/ r_e <r<r_i \}$,
in which we can simultaneously expand the external field when
$\varepsilon \to 0$ and keep its multipole moment structure. First of
all, from the fact that an arbitrary nonlinear coefficient
$h^{\mu\nu}_{{\rm can}(n)}$ in (\ref{eq:2.16}) is of order $O(2n, 2n+1, 2n)$
when $\varepsilon \to 0$ (see (5.5) in \cite{BD86}), we find that the
neglect of all nonlinear iterations with $n\geq 4$
permits the computation of the field up to the order $O(8,9,8)$ when
$\varepsilon \to 0$, i.e.
\begin{equation}
 h^{\mu\nu}_{\rm can} = G h^{\mu\nu}_{{\rm can}(1)}
  + G^2 h^{\mu\nu}_{{\rm can}(2)} + G^3 h^{\mu\nu}_{{\rm can}(3)} +
  O(8,9,8)\ . \label{eq:2.24}
\end{equation}
Secondly, it is shown in the appendix A that the second terms $q^{\mu\nu}
_{{\rm can}(2)}$ and $q^{\mu\nu}_{{\rm can}(3)}$ in the definitions of
the quadratic and cubic coefficients (\ref{eq:2.21}) have (at least)
the following orders of magnitude when $\varepsilon \to 0$:
\begin{mathletters}
\label{eq:2.25}
\begin{eqnarray}
 q^{\mu\nu}_{{\rm can}(2)} &=& O(7,7,7)\ , \label{eq:2.25a}\\
 q^{\mu\nu}_{{\rm can}(3)} &=& O(8,7,8)\ . \label{eq:2.25b}
\end{eqnarray}
\end{mathletters}
Therefore, from (\ref{eq:2.21}), we can write the expansion when
$\varepsilon \to 0$ of the canonical external field (\ref{eq:2.24}) in
the form
\begin{equation}
 h^{\mu\nu}_{\rm can} = Gh^{\mu\nu}_{{\rm can}(1)}
 + {\rm FP}_{B=0}\
 \Box^{-1}_R \left[ r^B (G^2\Lambda^{\mu\nu}_{{\rm can}(2)} + G^3
  \Lambda^{\mu\nu}_{{\rm can}(3)} ) \right] + O(7,7,7)\ , \label{eq:2.26}
\end{equation}
where the remainder term $O(7,7,7)$ is dominated by the contribution
(\ref{eq:2.25a}) of $q^{\mu\nu}_{\rm can(2)}$.

The source terms $\Lambda_{\rm can(2)}$ and $\Lambda_{\rm can(3)}$ in
(\ref{eq:2.26}) are given more explicitly by (\ref{eq:2.23}). To compute
their sum with an accuracy consistent with the remainder in (\ref{eq:2.26}),
we must know the quadratic metric $h_{\rm can(2)}$ up to the order
$O(6,5,6)$. This is of course analogous to the computation we have done
in \S~IIA, where we had first to solve Einstein's equations up to the
order $O(6,5,6)$ before reaching the looked-for order $O(8,7,8)$.  The
quadratic source at the order $O(6,5,6)$ is readily obtained by
substituting (\ref{eq:2.17})-(\ref{eq:2.18}) into (\ref{eq:2.23a}) and
discarding $O(6,5,6)$ terms.  We obtain
\begin{mathletters}
\label{eq:2.27}
\begin{eqnarray}
 G^2 \Lambda^{00}_{\rm can(2)} &=& - {14\over c^4}\, \partial_k V^{\rm ext}
 \partial_k V^{\rm ext} + O(6) \ , \label{eq:2.27a}\\
 G^2 \Lambda^{0i}_{\rm can(2)} &=& O (5)\ , \label{eq:2.27b} \\
 G^2 \Lambda^{ij}_{\rm can(2)} &=&  {4\over c^4}\, \left\{ \partial_i
  V^{\rm ext} \partial_j V^{\rm ext} - {1\over 2} \delta_{ij}
    \partial_k V^{\rm ext} \partial_k V^{\rm ext} \right\} + O(6) \ ,
     \label{eq:2.27c}
\end{eqnarray}
\end{mathletters}
from which we deduce
\begin{mathletters}
\label{eq:2.28}
\begin{eqnarray}
 G^2 h^{00}_{\rm can(2)} &=& -{7\over c^4}\,(V^{\rm ext})^2 + O(6)\ ,
  \label{eq:2.28a}\\
 G^2 h^{0i}_{\rm can(2)} &=& O(5)\ , \label{eq:2.28b}\\
 G^2 h^{ij}_{\rm can(2)} &=& -{4\over c^4}\,Z^{\rm ext}_{ij} + O(6)\ ,
  \label{eq:2.28c}
\end{eqnarray}
\end{mathletters}
where we have introduced the new external potential
\begin{equation}
 Z^{\rm ext}_{ij} = {\rm FP}_{B=0}\ \Box^{-1}_R
 \left[ r^B (-\partial_i V^{\rm ext} \partial_j V^{\rm ext} +
 {1\over 2} \delta_{ij} \partial_k V^{\rm ext} \partial_k V^{\rm ext})
 \right]\ . \label{eq:2.29}
\end{equation}
The justification of (\ref{eq:2.28}) is as follows. We know that the
post-Newtonian expansion of the regularized retarded operator
FP$\Box^{-1}_{R}$ acting on terms belonging to the quadratic source
$\Lambda^{\mu\nu}_{\rm can(2)}$ is equal to the expansion obtained from
the action of the regularized ``instantaneous" operator FP$I^{-1} :=
{\rm FP}\Sigma^\infty_{k=0}(\partial /c\partial t)^{2k} \Delta^{-k-1}$, where
$\Delta^{-k-1}$ is the $(k+1)$th iteration of the Poisson operator
$\Delta^{-1}$, modulo negligible terms of order $O(10,9,8)$ (see e.g.
(3.7) and (3.21), with $n=2$, in \cite{B93}). Thus, FP$\Box^{-1}_R$
acting on the remainder $O(6,5,6)$ in (\ref{eq:2.27}) is itself of order
$O(6,5,6)$. Furthermore, we know that FP$\Delta^{-1}$
(namely the first term in FP$I^{-1}$) gives simply $(U^{\rm ext})^2$ when
acting on $\Delta [(U^{\rm ext})^2]$, where $U^{\rm ext}$ is the
``Newtonian" potential associated with $V^{\rm ext}$ and given by
$U^{\rm ext} = G\Sigma {(-)^\ell\over \ell !} (\partial_L r^{-1}) M_L(t)$
(see e.g \cite{BD89} p.~395). Since $V^{\rm ext}
= U^{\rm ext} +O(2)$, we have FP$\Box^{-1}_R (\Delta [(V^{\rm ext})^2]) =
(V^{\rm ext})^2 +O(2)$ as has been used in (\ref{eq:2.28a}). Note that
this last fact implies that the trace $Z^{\rm ext} \equiv Z^{\rm ext}_{ii}$
of the potential (\ref{eq:2.29}) satisfies
\begin{equation}
 Z^{\rm ext} = {1\over 4} (V^{\rm ext})^2 + O(2)\ . \label{eq:2.30}
\end{equation}
Using now (\ref{eq:2.30}) and (\ref{eq:2.20}), we can write the
canonical field up to the order $O(6,5,6)$ in a form which is formally
identical to that of the inner field (\ref{eq:2.9}), i.e.
\begin{mathletters}
\label{eq:2.31}
\begin{eqnarray}
 h^{00}_{\rm can}&=& - {4\over c^2}\ V^{\rm ext} + {4\over c^4}\
 (W^{\rm ext} - 2 (V^{\rm ext})^2) + O(6)\ , \label{eq:2.31a} \\
 h^{0i}_{\rm can}&=& -{4\over c^3}\ V^{\rm ext}_i+ O(5)\ ,\label{eq:2.31b}\\
 h^{ij}_{\rm can}&=& -{4\over c^4}\ W^{\rm ext}_{ij}+O(6)\ ,\label{eq:2.31c}
\end{eqnarray}
\end{mathletters}
where we have defined
\begin{mathletters}
\label{eq:2.32}
\begin{equation}
 W^{\rm ext}_{ij} =V^{\rm ext}_{ij}+ Z^{\rm ext}_{ij}\ ,\label{eq:2.32a}\\
\end{equation}
whose trace is
\begin{equation}
 W^{\rm ext} \equiv W^{\rm ext}_{ii} = Z^{\rm ext} = {1\over 4}
   (V^{\rm ext})^2 + O(2)\ . \label{eq:2.32b}
\end{equation}
\end{mathletters}

 Finally, since the ``canonical" field (\ref{eq:2.31}) has the same form
in terms of the external potentials as the inner field (\ref{eq:2.9}) has
in terms of the inner potentials, it is clear that the source term $G^2
\Lambda^{\mu\nu}_{\rm can(2)} + G^3 \Lambda^{\mu\nu}_{\rm can(3)}$ in
the right-hand-side of (\ref{eq:2.26}) will be equal modulo the same
order $O(8,7,8)$ as in (\ref{eq:2.10}) to the truncated source
$\overline\Lambda^{\mu\nu}$ defined in (\ref{eq:2.12}), but computed
with the external potentials $V^{\rm ext}$, $V^{\rm ext}_i$ and $W^{\rm
ext}_{ij}$ instead of the inner potentials $V$, $V_i$, $W_{ij}$.
(Indeed, never a Laplace or d'Alembertian operator enters in the
effective gravitational source of Einstein's equations, which would give
a different result when acting on an external potential or on an inner
one.) Thus, we can write
\begin{equation}
 G^2 \Lambda^{\mu\nu}_{\rm can(2)} + G^3\Lambda^{\mu\nu}_{\rm can(3)}
 = \overline\Lambda^{\mu\nu} (V^{\rm ext}, V^{\rm ext}_i, W^{\rm ext}_{ij})
  + O(8,7,8)\ , \label{eq:2.33}
\end{equation}
where the right-hand-side is obtained from (\ref{eq:2.12}) by the simple
replacement $V$, $V_i$, $W_{ij} \to V^{\rm ext}, V^{\rm ext}_i, W^{\rm
ext}_{ij}$. Inserting (\ref{eq:2.33}) into (\ref{eq:2.26}), and
recalling that FP$\Box^{-1}_R$ acting on $O(8,7,8)$ is also $O(8,7,8)$,
yields our looked-for expression of the external field, namely
\begin{equation}
 h^{\mu\nu}_{\rm can} = Gh^{\mu\nu}_{\rm can(1)} + {\rm FP}_{B=0}
\ \Box^{-1}_R \left[ r^B \overline\Lambda^{\mu\nu}
(V^{\rm ext}, V^{\rm ext}_i, W^{\rm ext}_{ij})\right] + O(7,7,7)\ .
\label{eq:2.34}
\end{equation}
It is to be noticed that the remainder in (\ref{eq:2.34}) is $O(7,7,7)$
instead of the remainder $O(8,7,8)$ in the inner field (\ref{eq:2.14})
because it involves the not controlled contribution $O(7,7,7)$ of $q_{\rm
can(2)}^{\mu\nu}$ in (\ref{eq:2.25a}).  In the next section we match the
external field (\ref{eq:2.34}) to the corresponding inner field
(\ref{eq:2.14}).

\section{Matching of the internal and external fields}
\label{sec:3}

\subsection{Coordinate transformation between the internal and external
fields}
\label{ssec:3.1}

 We require that the internal field $h^{\alpha\beta}$ constructed in
\S~\ref{ssec:2.1} and the external field $h^{\mu\nu}_{\rm can}$
constructed in \S~\ref{ssec:2.2}  are {\it isometric} in their common domain
of validity, which is the exterior near-zone domain $D_i \cap D_e = \{
({\bf x},t) /\, r_e <r<r_i \}$ of the source. Denoting by $x^\alpha$ the
harmonic coordinates used in the inner domain $D_i$ (see \S~\ref{ssec:2.1}),
and by $x^\mu_{\rm can}$ the ``canonical" harmonic coordinates used in
the exterior domain $D_e$ (\S~\ref{ssec:2.2}), we thus look for a
compatible coordinate transformation
\begin{equation}
   x^\mu_{\rm can} (x) = x^\mu + \varphi^\mu (x)\ , \label{eq:3.1}
\end{equation}
where the vector $\varphi^\mu (x)$ is assumed to be in the form of a
multipolar and post-Newtonian expansion appropriate in $D_i \cap D_e$.
[In \S~\ref{ssec:2.2}, we have for notational convenience abusively denoted
by $x^\mu$ what really are the canonical harmonic coordinates
$x^\mu_{\rm can}$.] Since the two coordinate systems $x^\mu$ and
$x^\mu_{\rm can}$ are harmonic, the vector $\varphi^\mu$ satisfies the
(exact) relation
\begin{equation}
  \Box \varphi^\mu + h^{\alpha\beta} (x) \partial^2_{\alpha\beta}
    \varphi^\mu = 0\ , \label{eq:3.2}
\end{equation}
where $\Box = \eta^{\alpha\beta}\partial^2_{\alpha\beta}$. The (also exact)
transformation law of the field deviation $h^{\alpha\beta}$ under the
change of coordinates (\ref{eq:3.1}) is given by
\begin{equation}
 \eta^{\mu\nu} +h^{\mu\nu}_{\rm can} (x_{\rm can}) = {1\over |J|}
  (\delta^\mu_\alpha + \partial_\alpha \varphi^\mu)
  (\delta^\nu_\beta + \partial_\beta \varphi^\nu) [\eta^{\alpha\beta} +
  h^{\alpha\beta}(x)]\ , \label{eq:3.3}
\end{equation}
where $J=\det (\partial x_{\rm can} /\partial x)$ denotes the Jacobian
determinant of the coordinate transformation.

We now expand the transformation law (\ref{eq:3.3}) when $\varepsilon
\to 0$. The field deviation $h^{\alpha\beta}$ in the right-hand-side of
(\ref{eq:3.3}) is by (\ref{eq:2.9}) of order $O(2,3,4)$ when
$\varepsilon \to 0$. Furthermore, let us assume that the vector
$\varphi^\mu$ in the coordinate transformation (\ref{eq:3.1}) is of
order
\begin{equation}
  \varphi^\mu = O(3,4)\ . \label{eq:3.4}
\end{equation}
This assumption (which has already been made in previous works
\cite{BD89,DI91}) is proved below when we show that it leads to a consistent
matching. We can then easily see that the transformation law
(\ref{eq:3.3}) reduces up to the order $O(6,7,8)$ to a linear
transformation,
\begin{equation}
  h^{\mu\nu}_{\rm can} (x) = h^{\mu\nu} (x) +\partial\varphi^{\mu\nu} (x)
   + O(6,7,8)\ , \label{eq:3.5}
\end{equation}
where we have expressed both sides of the equation in terms of the inner
coordinates $x^\mu$, and where $\partial\varphi^{\mu\nu}$ denotes the
linear part of the coordinate transformation given by
\begin{equation}
 \partial\varphi^{\mu\nu} = \partial^\mu \varphi^\nu + \partial^\nu
 \varphi^\mu - \eta^{\mu\nu} \partial_\lambda \varphi^\lambda\ .
  \label{eq:3.6}
\end{equation}
By taking the divergence of (\ref{eq:3.6}) and using (\ref{eq:3.2}) we
obtain
\begin{mathletters}
\label{eq:3.7}
\begin{equation}
 \partial_\nu \partial\varphi^{\mu\nu} = \Box \varphi^\mu = -
h^{\rho\sigma} \partial_{\rho\sigma}^2 \varphi^\mu \ , \label{eq:3.7a}
\end{equation}
from which we deduce the order of magnitude
\begin{equation}
 \partial_\nu \partial\varphi^{\mu\nu} = O(7,8)\ . \label{eq:3.7b}
\end{equation}
\end{mathletters}
 Note that one could have a priori expected some nonlinear terms to
appear at the order $\varepsilon^6$ in the $ij$ component of the
equation (\ref{eq:3.5}) ---~for instance, terms like $h^{00}\partial^{(i}
\varphi^{j)}$ or $h^{0(i} \partial^{j)} \varphi^0$. Such nonlinear terms
are, however, absent at this order (as was also found in \cite{DI91}).
On the contrary, some nonlinear terms arise at the order $\varepsilon^6$
in the $00$ component of (\ref{eq:3.5}). These terms, which will be
needed in the following, are obtained by a short computation showing
that the $00$ component of (\ref{eq:3.5}), now valid up to the order
$O(8)$, is given by
\begin{equation}
 h^{00}_{\rm can}(x) = h^{00} (x) + \partial\varphi^{00}
  + 2h^{0\mu}\partial_\mu \varphi^0 - \partial_\mu(h^{00}\varphi^\mu)
  + \partial_i\varphi^0 \partial_i\varphi^0 +  O(8)\ . \label{eq:3.8}
\end{equation}
Here also we have expressed both sides of the equation in terms of the
inner coordinate system $x^\mu$.

\subsection{Matching of the compact-supported potentials $V$ and $V_i$}
\label{ssec:3.2}

A requisite in the matching procedure is to find the relations linking
the {\it multipole expansions} outside the source of the inner
potentials $V$, $V_i$ and $W_{ij}$ and the external potentials $V^{\rm
ext}$, $V_i^{\rm ext}$ and $W_{ij}^{\rm ext}$.  We deal in this
subsection with the case of the compact-supported potentials $V$ and
$V_i$.  The more complicated case of the non-compact-supported potential
$W_{ij}$ is reported in the next subsection.

 The inner and outer fields have been shown in \S~\ref{sec:2} to take,
up to the order $O(6,5,6)$, the same functional forms (\ref{eq:2.9}) and
(\ref{eq:2.31}) in terms of their respective potentials. On the other
hand, the coordinate transformation between them takes, up to this order
(and even up to a higher order), the linear form (\ref{eq:3.5}) with
(\ref{eq:3.6}). We first substitute into the sum of the $00$ component
and of the spatial trace $ii$ of (\ref{eq:3.5}) the inner and outer fields
(\ref{eq:2.9}) and (\ref{eq:2.31}). This leads to an equation whose
solution is easily seen to be
\begin{equation}
  V^{\rm ext} = V + c\, \partial_t \varphi^0 + O(4)\ . \label{eq:3.9}
\end{equation}
[Indeed, we have $\varphi^0 =O(3)$.] Similarly, by inserting
(\ref{eq:2.9}) and (\ref{eq:2.31}) into the $0i$ component of
(\ref{eq:3.5}), we obtain
\begin{equation}
  V^{\rm ext}_i = V_i - {c^3\over 4}\, \partial_i \varphi^0 + O(2)\ .
  \label{eq:3.10}
\end{equation}
Note that (\ref{eq:3.9}) is valid with the inclusion of relativistic
corrections $\varepsilon^2$, while (\ref{eq:3.10}) is valid at the
non-relativistic level only.

The relations (\ref{eq:3.9}) and (\ref{eq:3.10})  are {\it numerically}
true in the region $D_i \cap D_e$. We now transform them into {\it matching
equations}, i.e. equations relating mathematical expressions of the same
nature. To do this, we need only to replace the inner potentials in the
right-hand-sides of (\ref{eq:3.9}) and (\ref{eq:3.10}) by their multipole
expansions valid {\it outside} the source. This is simple because $V$ and
$V_i$ are the retarded integrals of the compact-supported mass and current
densities $\sigma$ and $\sigma_i$ (see (\ref{eq:2.3})). Thus the multipole
expansions ${\cal M} (V)$ and ${\cal M}(V_i)$ of $V$ and $V_i$ are given
by
\begin{mathletters}
\label{eq:3.11}
\begin{eqnarray}
 {\cal M} (V) &=& G \sum^{\infty}_{\ell =0} {(-)^\ell\over \ell !}
 \partial_L \left[ {1\over r} {\cal V}^L \left(t -{r\over c}\right)\right]
 \ , \label{eq:3.11a}\\
 {\cal M} (V_i) &=& G \sum^{\infty}_{\ell =0} {(-)^\ell\over \ell !}
 \partial_L \left[ {1\over r}{\cal V}^L_i \left(t-{r\over c}\right)\right]
 \ , \label{eq:3.11b}
\end{eqnarray}
\end{mathletters}
where the multipole moments ${\cal V}^L(t)$ and ${\cal V}^L_i(t)$ are given
by explicit integrals extending over the mass and current densities in the
source, namely
\begin{mathletters}
\label{eq:3.12}
\begin{eqnarray}
 {\cal V}^L(t) &=& \int d^3{\bf y} \hat y_L \int^1_{-1} dz\, \delta_\ell
 (z)\sigma \left({\bf y},t+z {|{\bf y}|\over c} \right)\ ,\label{eq:3.12a}\\
 {\cal V}^L_i(t) &=& \int d^3{\bf y} \hat y_L \int^1_{-1} dz\, \delta_\ell
 (z)\sigma_i \left({\bf y},t+z {|{\bf y}|\over c} \right)\ .\label{eq:3.12b}
\end{eqnarray}
\end{mathletters}
 In (\ref{eq:3.12}), $\hat y_L$ denotes the tracefree part of $y_L=y_{i_1}
y_{i_2}\cdots y_{i_\ell}$, and $\delta_\ell (z)$ is given by
\begin{equation}
 \delta_\ell (z) = {(2\ell +1)!!\over 2^{\ell+1}\ell !} (1-z^2)^\ell
\quad ; \qquad \int^1_{-1} dz\, \delta_\ell (z) =1\ . \label{eq:3.13}
\end{equation}
The formulas (\ref{eq:3.11})-(\ref{eq:3.13}) have been proved in the
appendix~B of \cite{BD89}. Now, we have $V={\cal M} (V)$ and $V_i ={\cal M}
(V_i)$ in $D_i \cap D_e$, hence we can write
\begin{eqnarray}
 V^{\rm ext} &=&{\cal M}(V) +c\,\partial_t\varphi^0 +O(4)\ ,\label{eq:3.14}\\
 V^{\rm ext}_i &=&{\cal M}(V_i) -{c^3\over 4}\,\partial_i\varphi^0
     +O(2)\ . \label{eq:3.15}
\end{eqnarray}
These matching equations relate the multipole expanded
external potentials $V^{\rm ext}$, $V^{\rm ext}_i$ given by
(\ref{eq:2.18}) to the multipole expansions ${\cal M}(V)$, ${\cal
M}(V_i)$ of the inner potentials, and can be used to obtain the
expressions of the ``canonical" moments $M_L(t)$ and $S_L(t)$ entering
(\ref{eq:2.18}) in terms of the source's parameters, with first
relativistic accuracy $\varepsilon^2$ in $M_L(t)$ and non-relativistic
accuracy in $S_L(t)$.  One can also obtain in this way the coordinate
change $\varphi^0$ (in the form of a multipole expansion).  This was the
method followed in \cite{BD89}, and we can check again that the results
of \cite{BD89} are indeed equivalent to (\ref{eq:3.14})-(\ref{eq:3.15}).
(We shall in particular recover below the results of \cite{BD89} by a
more general method.)

\subsection{Matching of the non-compact-supported potential $W_{ij}$}
\label{sec:3.3}

By inserting in the spatial components $ij$ of the transformation law
(\ref{eq:3.5}) the inner and outer fields (\ref{eq:2.9}) and
(\ref{eq:2.31}), we obtain
\begin{equation}
 W^{\rm ext}_{ij} = W_{ij} - {c^4\over 4} [ \partial_i \varphi^j +
 \partial_j \varphi^i - \delta_{ij} (\partial_0\varphi^0 + \partial_k
 \varphi^k) ] +O(2)\ . \label{eq:3.16}
\end{equation}
This equation is valid at the non-relativistic level only. Like
(\ref{eq:3.9}) and (\ref{eq:3.10}), it is numerically true in the
region $D_i \cap D_e$. To transform (\ref{eq:3.16}) into a matching
equation, we must first compute the multipole expansion, valid outside the
source, of the inner potential $W_{ij}$ which is, contrarily to the potentials
$V$ and $V_i$, of non-compact support.

Recall that $W^{\rm ext}_{ij}$ in the left-hand-side of (\ref{eq:3.16}) is
the sum of the potential $V^{\rm ext}_{ij}$ parametrizing the linear
metric (\ref{eq:2.17c}), and of the nonlinear potential $Z^{\rm ext}_{ij}
= Z_{ij} (V^{\rm ext})$ defined in (\ref{eq:2.29}). Now $V^{\rm ext}$ has
been matched to the multipole expansion of $V$ in the previous subsection:
$V^{\rm ext} ={\cal M}(V)+O(2)$ as deduced from (\ref{eq:3.14}).  Thus
we can write the external potential $W^{\rm ext}_{ij}$ as
\begin{equation}
 W^{\rm ext}_{ij} = V^{\rm ext}_{ij} + Z_{ij} ({\cal M}(V)) + O(2) \ ,
 \label{eq:3.17}
\end{equation}
where $Z_{ij} ({\cal M}(V))$ is given by
\begin{equation}
 Z_{ij} ({\cal M}(V)) = {\rm FP}_{B=0} \ \Box^{-1}_R
 \left[ r^B \left( - \partial_i {\cal M}(V)\partial_j {\cal M}(V)
 +{1\over 2}\,\delta_{ij}\partial_k {\cal M}(V) \partial_k {\cal M}(V)\right)
  \right]\ .  \label{eq:3.18}
\end{equation}
Note that it is crucial to replace $V^{\rm ext}$ in $Z_{ij} (V^{\rm ext})$
not by $V$ but by the {\it multipole} expansion ${\cal M}(V)$ of $V$ (modulo
$O(2)$). Indeed, contrarily to $V^{\rm ext} =V +O(2)$ which is true only in
the region $D_i \cap D_e$, the matching equation $V^{\rm ext} ={\cal M}(V) +
O(2)$ is an identity which is valid ``everywhere" and in particular on
the whole past null cone, issued from the considered field point, on
which depends the retarded integral in (\ref{eq:3.18}).

As for the inner
potential $W_{ij}$, we recall that it is given by
\begin{equation}
 W_{ij} = \Box^{-1}_R \left[ -4\pi G\, \sigma_{ij} - \partial_i V
 \partial_j V + {1\over 2} \delta_{ij} \partial_k V \partial_k V\right]\ .
  \label{eq:3.19}
\end{equation}
We shall now prove that the multipole expansion ${\cal M} (W_{ij})$,
valid outside
of source, of the (non-compact-supported) potential (\ref{eq:3.19}) reads
as
\begin{equation}
{\cal M}(W_{ij})= Z_{ij} ({\cal M}(V)) + G \sum^\infty_{\ell =0}\,
 {(-)^\ell\over \ell !}\,\partial_L \left[ {1\over r} {\cal W}^L_{ij}
\left( t-{r\over c}\right) \right]\ , \label{eq:3.20}
\end{equation}
where the first term is given by (\ref{eq:3.18}), and where the multipole
moments ${\cal W}^L_{ij} (t)$ in the second term are given by
\begin{equation}
 {\cal W}_{ij}^L(t) = {\rm FP}_{B=0} \int d^3 {\bf y}
  |{\bf y}|^B \hat y_L \int^1_{-1} dz\, \delta_\ell (z) \left[ \sigma_{ij}
 + {1\over 4\pi G} \left( \partial_i V\partial_j V-{1\over 2} \delta_{ij}
  \partial_k V\partial_k V\right)\right]
 \left({\bf y}, t+z {|{\bf y}|\over c}\right)\ ,
 \label{eq:3.21}
\end{equation}
where $\delta_\ell (z)$ is defined in (\ref{eq:3.13}).

The proof of (\ref{eq:3.20})-(\ref{eq:3.21}) goes as follows. The first
term in $W_{ij}$, involving the compact-supported matter stresses
$\sigma_{ij},$ can be
treated by formulas identical to (\ref{eq:3.11})-(\ref{eq:3.13}). We
thus focus the proof on the second term involving $\partial_i
V\partial_j V$ ---~from which we easily deduce the last term. Let us
consider the difference between (minus) this term and the corresponding
term in (\ref{eq:3.18}) involving the finite part at $B=0$  of the
retarded integral, i.e.
\begin{equation}
 X_{ij} = \Box^{-1}_R [\partial_i V\partial_j V] - {\rm FP}_
{B=0}\ \Box^{-1}_R [r^B \partial_i {\cal M} (V)
 \partial_j {\cal M} (V)]\ . \label{eq:3.22}
\end{equation}
The analytic continuation factor $r^B$ in the second retarded integral
deals with the singular behavior of the multipole expansions near the
spatial origin $r=0$.  The first integral does not need any
regularization factor because the integrand in perfectly regular at
$r=0$.  However, in order to make a better comparison between the two
integrals, let us introduce the same factor $r^B$, and the
``finite part" prescription, into the first integral.  Since this
integral is convergent, the ``finite part" prescription simply gives
back the value of the integral.  Hence we can rewrite (\ref{eq:3.22}) as
\begin{equation}
X_{ij} = {\rm FP}_{B=0}\ \Box^{-1}_R
 [ r^B ( \partial_i V\partial_j V - \partial_i {\cal M} (V)
 \partial_j {\cal M} (V) )]\ . \label{eq:3.24}
\end{equation}
The important point is that, under the form (\ref{eq:3.24}), we see that
$X_{ij}$ {\it does} have a compact support limited to the
distribution of matter in the source.  Indeed, outside the compact
support of the source (i.e., in the domain $D_e$), the potential $V$
numerically agrees with its multipole expansion ${\cal M}(V)$ and
therefore the integrand in (\ref{eq:3.24}) is identically zero.  (This
is at this point that we need to assume that the multipole expansions in
the ``canonical" construction of the exterior metric in $D_e$ involve an
infinite number of multipoles.) We can thus compute the multipole
expansion ${\cal M} (X_{ij})$ of (\ref{eq:3.24}) in $D_e$ by exactly the
same formulas (\ref{eq:3.11})-(\ref{eq:3.13}) as was used for the
compact-supported potentials $V$ and $V_i$. The only difference is that the
analytic continuation factor $r^B$ must be kept inside the integral, and
that one must apply to the formulas the ``finite part" prescription.
One is thus led to the multipole expansion
\begin{equation}
 {\cal M} (X_{ij}) = \sum^\infty_{\ell =0} {(-)^\ell\over \ell !}\,
  \partial_L \left[ {1\over r} {\cal X}^L_{ij} \left( t-{r\over c}
  \right)\right]\ , \label{eq:3.25}
\end{equation}
where the multipole moments ${\cal X}^L_{ij} (t)$ are given by
\begin{equation}
 {\cal X}_{ij}^L(t) = {-1\over 4\pi}\
 {\rm FP}_{B=0} \int d^3 {\bf y}
  |{\bf y}|^B \hat y_L \int^1_{-1} dz\, \delta_\ell (z) \left[
 \partial_i V\partial_j V - \partial_i {\cal M}(V) \partial_j {\cal M}(V)
  \right] \left({\bf y}, t+z {|{\bf y}|\over c}\right)\ .
 \label{eq:3.26}
\end{equation}
[The ``finite part" commutes with the derivation operator $\partial_L$
in (\ref{eq:3.25}).] We now prove that the second term in (\ref{eq:3.26}),
which involves the multipole expansion ${\cal M}(V)$ in the integrand,
is in fact {\it zero} by analytic continuation. Indeed, the multipole
expansion ${\cal M}(V) ({\bf y},t)$ is by (\ref{eq:3.11a}) an expansion
of the type $\Sigma \hat n_{L'} |{\bf y}|^{-p} {\cal F} (t-|{\bf y}|/c)$,
where $L'$ is some multi-index, $p$ is some integer and ${\cal F}$ some
function of time (all indices suppressed). Thus, ${\cal M}(V) ({\bf y},t+z
|{\bf y}|/c)$ is of the type $\Sigma \hat n_{L'} |{\bf y}|^{-p} {\cal F}
(t+(z-1)|{\bf y}|/c)$. By differentiating and squaring the latter expansion,
we see that the product $\partial_i {\cal M}(V)\partial_j {\cal M}(V)
({\bf y},t+z|{\bf y}|/c)$ entering (\ref{eq:3.26}) is an expansion of the
same type $\Sigma\hat n_{L''} |{\bf y}|^{-q} {\cal G} (t+(z-1)|{\bf y}|/c)$,
with $L''$ a multi-index, $q$ an integer and ${\cal G}$ a function of time.
We then expand by Taylor's formula each function ${\cal G}$ when $c\to
+\infty$. This introduces many powers of $|{\bf y}|$ and of $(z-1)$.
Multiplying by $\delta_\ell (z)$ and integrating over $z$, we find a
multipole expansion of the type $\Sigma \hat n_{L''} |{\bf y}|^k {\cal H}
(t)$, where $k$  is an integer and ${\cal H}$ another function of
time;  then multiplying by $\int d^3{\bf y} |{\bf y}|^B \hat y_L$ and
performing
the angular integration (using $d^3{\bf y} = d|{\bf y}| |{\bf y}|^2
d\Omega$ and $\int d\Omega \hat n_{L''} \hat n_L = {\rm const.} \
\delta_{\ell,\ell''})$, we arrive at a multipole expansion of the type
$\Sigma \hat n_L {\cal H} (t) \int^{+\infty}_0 d |{\bf y}| |{\bf
y}|^{B+m}$, where $m$ is an integer.  Finally each of the latter
integrals $\int^{+\infty}_0 d |{\bf y}| |{\bf y}|^{B+m}$ is zero by
analytic continuation.  [Indeed, we cut the integrals into two pieces,
$I_1 = \int^Y_0 d|{\bf y}| |{\bf y}|^{B+m}$ and $I_2 = \int^{+\infty}_Y
d|{\bf y}| |{\bf y}|^{B+m}$, where $Y$ is some constant $>0$.  By
choosing the real part of $B$ to be such that $Re\, B +m>-1$, we compute
$I_1 =Y^{B+m+1}/(B+m+1)$;  and by choosing the real part of $B$ to be
such that $Re\, B+m<-1$, we compute $I_2 =-Y^{B+m+1}/(B+m+1)$. Then both
$I_1$ and $I_2$ can be analytically continued for all complex values of
$B$ except the single value $-m-1$.  The integral $\int^{+\infty}_0
d|{\bf y}| |{\bf y}|^{B+m}$ is the sum of the analytic continuations of
$I_1$ and $I_2$, and is thus identically zero on the whole complex plane
(including $-m-1$).] Note that the proof that the second term in
(\ref{eq:3.26}) is zero can be easily extended to the case where we
have, instead of $\partial_i {\cal M}(V) \partial_j {\cal M}(V)$, an
arbitrary nonlinear multipolar product of the type $\partial_{\mu} {\cal M}_1
\partial_{\nu}
{\cal M}_2 {\cal M}_3\cdots$, where ${\cal M}_1 ({\bf y},t)$, ${\cal M}_2
({\bf y},t)$, ${\cal M}_3 ({\bf y},t)\cdots$ denote formal multipolar
and post-Newtonian expansions of
the type $\Sigma \hat n_{L'} |{\bf y}|^p (\ln |{\bf y}|)^q {\cal K}
(t)$, where $p,q$ are integers (such expansions are known to arise in
higher-order nonlinear approximations of the external field
\cite{BD86}).  Thus, the multipole moments ${\cal X}^L_{ij}(t)$ in
(\ref{eq:3.26}) can be simply written as
\begin{equation}
 {\cal X}^L_{ij} (t) = {-1\over 4\pi}\ {\rm FP}_{B=0}
  \int d^3 {\bf y} |{\bf y}|^B \hat y_L
  \int^1_{-1} dz\delta_\ell (z)(\partial_iV\partial_jV) \left( {\bf y},t+z
   {|{\bf y}|\over c} \right)\ , \label{eq:3.27}
\end{equation}
and the expressions (\ref{eq:3.20})-(\ref{eq:3.21}) are then easily
deduced.  Indeed, we write $W_{ij}$ as the sum of $Z_{ij} ({\cal
M}(V))$, of the compact-support integral $\Box^{-1}_R [-4\pi\,
G\,\sigma_{ij}]$ and of the term $-X_{ij} +{1\over 2}\, \delta_{ij}
X_{kk}$.  The multipole expansion ${\cal M} (W_{ij})$ is thus the sum of
$Z_{ij}({\cal M}(V))$, which is already a multipole expansion, of the
multipole expansion of $\Box^{-1}_R [-4\pi G\sigma_{ij}]$ computed by
formulas like (\ref{eq:3.11})-(\ref{eq:3.12}), and of $-{\cal M}(X_{ij})
+ {1\over 2} \delta_{ij} {\cal M}(X_{kk})$ as deduced from
(\ref{eq:3.25}) and (\ref{eq:3.27}).  We add the factor $|{\bf y}|^B$
and the ``finite part" prescription into the moments of the expansion of
$\Box^{-1}_R [-4\pi G\sigma_{ij}]$ (which are convergent anyway), and
the result is ${\cal M}(W_{ij})$ given by
(\ref{eq:3.20})-(\ref{eq:3.21}).

 One should note the remarkable role of the analytic continuation factor
$|{\bf y}|^B$ in (\ref{eq:3.26}) and (\ref{eq:3.27}). The integral
(\ref{eq:3.26}) is of compact support and is thus perfectly well-defined
at infinity $|{\bf y}|\to +\infty$. The role of the factor $|{\bf y}|^B$
in (\ref{eq:3.26}) is to deal with the singular behavior at the {\it
origin} $|{\bf y}| =0$ of the second term in the integrand, involving
the multipole expansion ${\cal M}(V)$.  That is, one can compute
(\ref{eq:3.26}) by choosing $Re\,B$ to be a large {\it positive} number
so as to kill the bad behavior of the integral at the origin, then
analytically continuing the integral near $B=0$ and deducing the finite
part at $B=0$. On the contrary, the integral (\ref{eq:3.27}) is perfectly
well-defined near the origin $|{\bf y}|=0$, but it is not apparently of
compact support.  The factor $|{\bf y}|^B$ in (\ref{eq:3.27}) then deals
with the a priori bad behavior of the integral at {\it infinity} $|{\bf
y}|\to \infty$, where $\hat y_L \partial_i V \partial_j V$ behaves for
large $\ell$ like a large power $\sim |{\bf y}|^{\ell -4}$ blowing up at
infinity.  That is, one can compute (\ref{eq:3.27}) by choosing $Re\,B$
to be a large {\it negative} number so as to make the integral convergent
when $|{\bf y}|\to \infty$. These two different procedures,
$Re\,B$ a large positive number in (\ref{eq:3.26}) and a large negative
number in (\ref{eq:3.27}), give the same numerical result, as we have
just proved. In conclusion, the form (\ref{eq:3.27}) although not
apparently of compact support is however, thanks to the properties of
analytic continuation, numerically equal to the compact-supported form
(\ref{eq:3.26}).  It is evident that the form
(\ref{eq:3.27}), where the analytic continuation factor deals with the
behavior of the integral at {\it infinity} from the source, is the one
which should be used in applications.

The equation (\ref{eq:3.16}) can finally be replaced by the matching
equation
\begin{equation}
 W^{\rm ext}_{ij} = {\cal M}(W_{ij}) - {c^4\over 4} [ \partial_i \varphi^j +
 \partial_j \varphi^i - \delta_{ij} (\partial_0\varphi^0 + \partial_k
 \varphi^k) ] +O(2)\ , \label{eq:3.28}
\end{equation}
where the multipole expansion ${\cal M}(W_{ij})$ is given by
(\ref{eq:3.20})-(\ref{eq:3.21}). This equation can be equivalently rewritten
as
\begin{equation}
 V^{\rm ext}_{ij} = G \sum^\infty_{\ell =0}\,
 {(-)^\ell\over \ell !}\,\partial_L \left[ {1\over r} {\cal W}^L_{ij}
 \left( t-{r\over c}\right) \right] - {c^4\over 4} [ \partial_i \varphi^j +
 \partial_j \varphi^i - \delta_{ij} (\partial_0\varphi^0 + \partial_k
 \varphi^k) ] +O(2)\ , \label{eq:3.29}
\end{equation}
and in this form can be used to compute, if desired, the vector
$\varphi^i$ (since $\varphi^0$ is known from (\ref{eq:3.15})). To this
end, one needs to decompose the moments ${\cal W}^L_{ij}$ into
irreducible STF tensors with respect to the indices $ij$ and $L$.

\subsection{Matching equation at the post-Newtonian order $\varepsilon^6$}
\label{ssec:3.4}

We now have in hand all the material needed to match the inner metric
(\ref{eq:2.14}), namely
\begin{equation}
 h^{\alpha\beta} (x) =\Box^{-1}_R \left[ {16\pi G\over c^4}\
\overline\lambda (V,W) T^{\alpha\beta} + \overline\Lambda^{\alpha\beta}
(V, V_i, W_{ij}) \right] + O(8,7,8)\ , \label{eq:3.30}
\end{equation}
where $\overline\lambda$ and $\overline\Lambda^{\alpha\beta}$ are
defined in (\ref{eq:2.11}) and (\ref{eq:2.12}), to the corresponding
outer metric (\ref{eq:2.34}), namely
\begin{equation}
h^{\mu\nu}_{\rm can} (x_{\rm can}) = Gh^{\mu\nu}_{\rm can(1)}(x_{\rm can})
  + {\rm FP}_{B=0}\ \Box^{-1}_R \left[ r^B
  \overline\Lambda^{\mu\nu} (V^{\rm ext}, V_i^{\rm ext}, W_{ij}^{\rm ext})
  \right] + O(7,7,7)\ , \label{eq:3.31}
\end{equation}
where $h^{\mu\nu}_{\rm can(1)}$ is the linear metric
(\ref{eq:2.17})-(\ref{eq:2.18})  and $\overline\Lambda^{\mu\nu}$ is the
same expression as in (\ref{eq:3.30}) but expressed in terms of the
external potentials $V^{\rm ext}$, $V^{\rm ext}_i$ and $W^{\rm
ext}_{ij}$. The metrics (\ref{eq:3.30}) and (\ref{eq:3.31}) are given
in their respective coordinate systems $x^\mu$ and $x^\mu_{\rm can}$, and
are valid respectively in $D_i$ and $D_i \cap D_e$.

In the two previous subsections, we have related the external potentials
$V^{\rm ext}$, $V^{\rm ext}_i$ and $W^{\rm ext}_{ij}$ to the multipole
expansions ${\cal M}(V)$, ${\cal M}(V_i)$ and ${\cal M}(W_{ij})$ of the
internal potentials $V$, $V_i$ and $W_{ij}$. These relations, which are
(\ref{eq:3.14})-(\ref{eq:3.15}) and (\ref{eq:3.28}), allow us to compute
the effective nonlinear source $\overline\Lambda^{\mu\nu} (V^{\rm ext},
V_i^{\rm ext}, W^{\rm ext}_{ij}$) in the right-hand-side of the outer
metric (\ref{eq:3.31}) in terms of the inner potentials.  Indeed, by
using (\ref{eq:3.14})-(\ref{eq:3.15}) and (\ref{eq:3.28}) into the
expression (\ref{eq:2.12}) of $\overline\Lambda^{\mu\nu}$, and by using
$\Box {\cal M}(V) = \Box {\cal M} (V_i) =0$ and $\Box \varphi^\mu =
O(7,8)$ (see (\ref{eq:3.7b})), we find
\begin{equation}
\overline{\Lambda}^{\mu\nu} (V^{\rm ext}, V^{\rm ext}_i, W^{\rm ext}_{ij})
 =\overline{\Lambda}^{\mu\nu} ({\cal M}(V),{\cal M}(V_i),{\cal M}(W_{ij})) +
 \Box \Omega^{\mu\nu} + O(8,7,8)\ , \label{eq:3.32}
\end{equation}
where in the right-hand-side extra terms appear which are due to the
coordinate transformation between the inner and outer metrics, and which
can be written as a d'Alembertian operator $\Box$ acting on the tensor
\begin{mathletters}
\label{eq:3.33}
\begin{eqnarray}
\Omega^{00} &=& - {8\over c^3}\, \left[ {\cal M}(V) \partial_t \varphi^0
  + {\cal M}(V_i) \partial_i\varphi^0 - {c\over 2} \partial_\mu
 ({\cal M} (V) \varphi^\mu)\right] + \partial_i \varphi^0
   \partial_i\varphi^0\ , \label{eq:3.33a}\\
\Omega^{0i} &=& 0\ , \label{eq:3.33b}\\
\Omega^{ij} &=& 0\ . \label{eq:3.33c}
\end{eqnarray}
\end{mathletters}
[The only non-zero component is $\Omega^{00}$ which is of order $O(6)$.]
Now, to evaluate (\ref{eq:3.31}) we must apply on both sides of
(\ref{eq:3.32}) the regularized inverse d'Alembertian operator
FP$\Box^{-1}_R$. One easily checks that FP$\Box^{-1}_R (\Box\Omega^{00})
=\Omega^{00} + O(8)$.  Indeed, this follows from the structure of
$\Omega^{00}$ which is made at leading level $\varepsilon^6$ of terms of
the type $\Sigma \hat n_L r^{-(\ell+2+2k)}$, where $k$ is a positive
integer (see \cite{BD89} p.~395). [${\cal M}(V), {\cal M}(V_i)$ and
$\varphi^0$, $\varphi^i$ all have the structure $\Sigma \hat n_{L'}
r^{-(\ell '+1)}$ at leading level.] Hence, we find that the outer metric
(\ref{eq:3.31}) reads as
\begin{equation}
 h^{\mu\nu}_{\rm can} (x_{\rm can}) = Gh^{\mu\nu}_{\rm can(1)} (x_{\rm can})
 + {\rm FP}_{B=0}\ \Box^{-1}_R \left[ r^B
    \overline \Lambda^{\mu\nu} ({\cal M}(V), {\cal M}(V_i), {\cal M}(W_{ij}
     )) \right] +\Omega^{\mu\nu}+O(7,7,7)\ . \label{eq:3.34}
\end{equation}
On the other hand, we have shown in \S~\ref{ssec:3.1} that the inner and
outer metrics should be linked by a coordinate transformation involving
nonlinear terms at order $\varepsilon^6$ in its 00 component, see
(\ref{eq:3.8}). Using $h^{00}=- {4\over c^2} {\cal M} (V)+\cdots$ and
$h^{0i}= -{4\over c^3} {\cal M} (V_i) +\cdots$ as is appropriate in $D_i
\cap D_e$, it is easy to see that the nonlinear terms in (\ref{eq:3.8})
are precisely equal to $\Omega^{00}$ given by (\ref{eq:3.33a}), modulo
$O(8)$.  Hence, the coordinate transformation
(\ref{eq:3.5})-(\ref{eq:3.8}) is in fact given by
\begin{equation}
 h^{\mu\nu}_{\rm can} (x) = h^{\mu\nu} (x) + \partial\varphi^{\mu\nu} +
 \Omega^{\mu\nu} + O(8,7,8)\ , \label{eq:3.35}
\end{equation}
(where the linear part $\partial\varphi^{\mu\nu}$ of the coordinate
transformation is defined by (\ref{eq:3.6}),
and where both sides are expressed in terms of the inner coordinates
$x^\mu$). Substituting in the left-hand-side of (\ref{eq:3.35}) the outer
metric (\ref{eq:3.34}), and in the right-hand-side the inner metric
(\ref{eq:3.30}), we arrive at the following equation for the external
linear metric (in the coordinates $x^\mu$):
\begin{eqnarray}
 Gh^{\mu\nu}_{\rm can(1)} (x) =&& \Box^{-1}_R \left[ {16\pi G\over c^4}
 \overline\lambda (V,W) T^{\mu\nu} +\overline\Lambda ^{\mu\nu} (V,V_i,
   W_{ij}) \right] \nonumber \\
 &&- {\rm FP}_{B=0}\ \Box^{-1}_R [r^B \overline\Lambda
  ^{\mu\nu} ({\cal M} (V), {\cal M} (V_i), {\cal M}(W_{ij}))]
  + \partial \varphi^{\mu\nu} + O(7,7,7)\ . \label{eq:3.36}
\end{eqnarray}
The nonlinear part $\Omega^{\mu\nu}$ of the coordinate transformation
has been cancelled, and it remains only the linear coordinate
transformation $\partial\varphi^{\mu\nu}$.

The equation (\ref{eq:3.36}) is numerically valid in $D_i \cap D_e$, but
must now be transformed into a matching equation. The reasoning is
exactly the same as the one we followed in dealing with the
non-compact-supported potential $W_{ij}$ in \S~\ref{sec:3.3}, and we
simply repeat here the arguments. First, the term in (\ref{eq:3.36})
involving the compact-supported matter stress-energy tensor $T^{\mu\nu}$
is treated by formulas such as (\ref{eq:3.11})-(\ref{eq:3.13}). Then
consider, analogously to (\ref{eq:3.22}), the difference between the
two terms involving $\overline\Lambda^{\mu\nu}$ in (\ref{eq:3.36}). By
adding the factor $r^B$ and the ``finite part" prescription in the first
of these terms (whose integrand is regular at $r=0$) we can write this
difference, analogously to (\ref{eq:3.24}), in a manifestly
compact-supported form.  (Indeed, the potentials $V$, $V_i$ and $W_{ij}$
are numerically equal in $D_e$ to their multipole expansions ${\cal
M}(V)$, ${\cal M} (V_i)$ and ${\cal M}(W_{ij})$.) Hence, we can also
apply to this difference the formulas (\ref{eq:3.11})-(\ref{eq:3.13})
and obtain, analogously to (\ref{eq:3.25})-(\ref{eq:3.26}), some
expressions for the multipole moments allowing for the analytic
continuation factor $|{\bf y}|^B$ and the finite part at $B=0$.  Finally
the contribution associated with the multipole expanded source
$\overline\Lambda^{\mu\nu} ({\cal M}(V), {\cal M} (V_i), {\cal M}(W_{ij}))$
in these multipole moments is shown, analogously to (\ref{eq:3.27}), to
be zero by analytic continuation.  This follows from the fact that
$\overline\Lambda^{\mu\nu} ({\cal M}(V), {\cal M} (V_i), {\cal
M}(W_{ij}))$ is a sum of quadratic or cubic terms $\partial_{\rho} {\cal M}_1
\partial_{\sigma} {\cal M}_2$ or $\partial_{\rho} {\cal M}_1\partial_{\sigma}
{\cal M}_2
{\cal M}_3$ where the ${\cal M}_k$'s admit formal multipolar and post-Newtonian
expansions of the type $\Sigma \hat n_{L'}
|{\bf y}|^p (\ln |{\bf y}|)^q {\cal K}(t)$, with $p,q$ some integers $(q=0$ in
the case of ${\cal M}(V)$ and ${\cal M} (V_i)$, but $q=0$ or 1 in the
case of ${\cal M}(W_{ij}))$;  see the discussion above (\ref{eq:3.27}).
An important result of this paper can now be written down.  We introduce
an effective (truncated) {\it total stress-energy tensor} of the matter
fields and of the gravitational field,
\begin{equation}
 \overline\tau^{\mu\nu}(V,V_i,W_{ij}) = \overline\lambda (V,W) T^{\mu\nu}
  + {c^4\over 16\pi G}\overline\Lambda^{\mu\nu} (V,V_i, W_{ij}) \ ,
    \label{eq:3.37}
\end{equation}
where $\overline\lambda$ and $\overline\Lambda^{\mu\nu}$ are given by
(\ref{eq:2.11}) and (\ref{eq:2.12}). By (\ref{eq:2.13}) this tensor satisfies
\begin{equation}
 \partial_\nu \overline\tau^{\mu\nu} = O(3,4)\  \ . \label{eq:3.38}
\end{equation}
Then we can transform (\ref{eq:3.36}) into the matching
equation
\begin{equation}
 Gh^{\mu\nu}_{\rm can(1)} [M_L,S_L] = -{4G\over c^4} \sum^\infty_{\ell =0}
 {(-)^\ell\over \ell !} \partial_L \left[ {1\over r}\,
  \overline{\cal T}^{\mu\nu}_L \left( t-{r\over c}\right)\right] +
  \partial\varphi^{\mu\nu} + O(7,7,7)\ , \label{eq:3.39}
\end{equation}
relating the exterior linear metric (\ref{eq:2.17})-(\ref{eq:2.18}), that
we recall is a functional of the ``canonical" moments $M_L$ and $S_L$,
to the total truncated stress-energy tensor (\ref{eq:3.37}) via the
multipole moments
\begin{equation}
 \overline{\cal T}^{\mu\nu}_L(t) = {\rm FP}_{B=0} \int
  d^3 {\bf y}|{\bf y}|^B \hat y_L \int^1_{-1} dz\, \delta_\ell (z)
  \overline\tau^{\mu\nu} \left( {\bf y}, t+z {|{\bf y}|\over c}\right)\ .
  \label{eq:3.40}
\end{equation}

 The result (\ref{eq:3.39})-(\ref{eq:3.40}) is especially simple. It says
that the {\it linear} metric $h^{\mu\nu}_{\rm can(1)}$ is equal, modulo
the {\it linear} coordinate transformation $\partial\varphi^{\mu\nu}$
(and modulo $O(7,7,7)$ terms), to the multipole expansion outside the
source we would obtain {\it if} the effective total stress-energy tensor
$\overline\tau^{\mu\nu}$ had a compact support limited to the material
source only.  The difference is that the moments (\ref{eq:3.40}) carry
the analytic continuation factor $|{\bf y}|^B$ and the ``finite part" at
$B=0$ to deal with the poles at $B=0$ coming from the behavior of the
integral at its upper bound $|{\bf y}| \to +\infty$.  When no poles
arise as will be the case at the 2-PN approximation (see \S~\ref{sec:4}),
we can say that (\ref{eq:3.39})-(\ref{eq:3.40}) justifies the formal
procedure followed by Epstein and Wagoner \cite{EW75} and Thorne
\cite{Th80} to compute the multipole moments.  (However recall that in
the works \cite{EW75} and \cite{Th80} these multipole moments are
assumed to be the moments which are radiated at infinity, while one has
still to add to these moments all tails and nonlinear contributions in
the radiation field, see \S~\ref{sec:4} below.) Let us emphasize again
that the expression (\ref{eq:3.40}) is not manifestly of compact-supported
form but is numerically equal to the expression obtained by multipole
expanding the compact-supported right-hand-side of (\ref{eq:3.36}).  In
particular, this shows that the multipole moments (\ref{eq:3.40}) are
{\it retarded} functionals of the source's parameters, depending on the
source at times $t'\leq t$ only (or $t' \leq t-r/c$ in (\ref{eq:3.39})),
contrarily to what is apparent on their expressions (\ref{eq:3.40}) but
in accordance with what they must be.

\section{generation of gravitational waves}
\label{sec:4}

\subsection{Relations between the canonical moments and the source
moments}
\label{ssec:4.1}

In order to find from the matching equation (\ref{eq:3.39})-(\ref{eq:3.40})
the expressions of the ``canonical'' moments $M_L$, $S_L$ as functions
of the source's parameters, it suffices simply to decompose the
reducible moments $\overline{\cal T}^{\mu\nu}_L (t)$ into irreducible
STF multipole moments.  This has already been done by Damour and Iyer
\cite{DI91'} in the case of {\it linearized} gravity, i.e.  in the case
where the total stress-energy tensor $\overline\tau^{\mu\nu}$ in
(\ref{eq:3.39}) is replaced by the usual compact-supported stress-energy
tensor $T^{\mu\nu}$ of the matter fields, supposed to be exactly
conserved:  $\partial_\nu T^{\mu\nu}=0$, and where of course there are no
analytic continuation factors in the expressions of the moments.

 Let us prove that the computation done in \cite{DI91'} basically applies
to our case, which involves both analytic continuation factors and a
stress-energy tensor $\overline\tau^{\mu\nu}$ which is only approximately
conserved:  $\partial_\nu \overline\tau^{\mu\nu}=0 (3,4)$ by
(\ref{eq:3.38}). To this end, we need only to check that the multipole
expansion in the right-hand-side of (\ref{eq:3.39}) is divergenceless up
to $O(3,4)$ as a consequence of the approximate conservation of
$\overline\tau^{\mu\nu}$.  We take the time derivative of the $0\mu$
components of the multipole moments (\ref{eq:3.40}), use $c^{-1}
\partial_t \overline\tau^{0\mu} = -\partial_j \overline\tau^{j\mu} +
O(3,4)$, perform some integrations by parts both with respect to ${\bf
y}$ and to $z$, and finally insert the identity $(d/dz)^2
\delta_{\ell+1}(z) =(2\ell+1)(2\ell+3)[\delta_{\ell-1} (z) -
\delta_\ell (z)]$.  These operations result in
\begin{eqnarray}
 {d\over cdt} \overline{\cal T}^{0\mu}_L(t) =&& \ell\ \overline{\cal T}
 ^{\mu\langle i_\ell}_{L-1\rangle} (t) + {1\over 2\ell +3}
 \left( {d\over cdt}\right)^2 \overline{\cal T}^{j\mu}_{jL} (t) \nonumber \\
 &&+ {\rm FP}_{B=0} \left\{ B \int d^3{\bf y}
    |{\bf y}|^{B-2} y_j \hat y_L \int^1_{-1} dz\delta_\ell (z)
   \overline\tau^{j\mu} ({\bf y},t+z|{\bf y}|/c) \right\}\nonumber \\
 &&+ O(3,4)\ . \label{eq:4.1}
\end{eqnarray}
The first two terms in the right-hand-side of (\ref{eq:4.1}) are the
ones which would ensure the exact zero divergency of the multipole
expansion in (\ref{eq:3.39}). (${\cal T}^{\mu\langle i_\ell}_{L-1\rangle}$
means the STF part of ${\cal T}^{\mu i_\ell}_{L-1}$.) The third term is the
finite part at $B=0$ of an integral having explicitly $B$ as a factor.
This term arises because of the derivation of the factor $|{\bf y}|^B$
during the integration by parts.  (Note that we have discarded a surface
term which is easily seen to be zero by analytic continuation.) Finally,
the remainder $O(3,4)$ comes from the only approximate conservation
of $\overline\tau^{\mu\nu}$.  We now show that the third term in
(\ref{eq:4.1}) in fact vanishes up to order $O(3,4)$, which means,
thanks to the explicit factor $B$, that the integral has at this order
no simple pole at $B=0$.  This follows from a
general lemma which we now state and which will turn out to be very
useful in the remaining of the paper.  Let ${\bf y}$ be a ``source"
point, ${\bf x}_1,$ ${\bf x}_2$,\dots, ${\bf x}_{\rm n}$ be n ``field"
points, and $\alpha_0$, $\alpha_1$,\dots,$\alpha_{\rm n}$ be ${\rm n}+1$
real numbers.  Then we can state that
\begin{equation}
 {\rm FP}_{B=0} \left\{ B \int d^3 {\bf y}
  |{\bf y}|^{B+\alpha_0} \hat y_L |{\bf y}-{\bf x}_1|^{\alpha_1} \cdots
  |{\bf y}-{\bf x}_{\rm n}|^{\alpha_{\rm n}} \right\} = 0\label{eq:4.2}
\end{equation}
if the sum $\alpha_0 +\alpha_1 +\cdots +\alpha_{\rm n}$ is {\it not} an
odd (positive or negative) integer.  To prove that this is true, we need
to investigate the behavior of the integrand at the bound $|{\bf y}|\to
+\infty$ of the integral, where each $|{\bf y}-{\bf x}_1|^{\alpha_1},$\dots~
admits a multipole expansion of the type $\Sigma x^{L_1-1} \partial_{L_1}
|{\bf y}|^{\alpha_1}\cdots$ (coefficients suppressed).  Thus, one is led
to consider the behavior of integrals of the type $\int d^3{\bf y}|{\bf
y}|^{B+\alpha_0} \hat y_L \partial_{L_1} |{\bf y}|^{\alpha_1}\cdots
\partial_{L_{\rm n}}|{\bf y}|^{\alpha_{\rm n}}$.  By performing the
integration over the angles (using $d^3{\bf y}=|{\bf y}|^2 d|{\bf
y}|d\Omega$), we find that these integrals are zero if $\ell +\ell_1
+\cdots +\ell_{\rm n}$ is an odd integer.  When $\ell+\ell_1 +\cdots
+\ell_{\rm n}$ is an even integer, a single type of radial integral
remains, which is $\int d|{\bf y}||{\bf y}|^{B+\beta}$ where $\beta
=2+\alpha_0 +\ell +(\alpha_1 -\ell_1) +\cdots +(\alpha_{\rm n}
-\ell_{\rm n})$.  A pole at $B=0$ will arise only if $\beta=-1$, and
thus, since $\ell +\ell_1+\cdots +\ell_{\rm n}$ is an even integer, only
if $\alpha_0 + \alpha_1 +\cdots +\alpha_{\rm n}$ is an odd integer.  If
$\alpha_0 +\alpha_1 + \cdots +\alpha_{\rm n}$ is not an odd integer,
there is no pole and (\ref{eq:4.2}) is true by virtue of the explicit
factor $B$ in front of the integral.  (Note that this condition for
(\ref{eq:4.2}) to hold is sufficient but by no means necessary.) Now, a
non-zero contribution to the third term in (\ref{eq:4.1}) can come only
from the non-compact supported part of $\overline\tau^{j\mu}$ in the
integrand, i.e.  the part involving $\overline\Lambda^{j\mu}$ (see
(\ref{eq:3.37})).  We first expand the argument $t+z |{\bf y}|/c$ when
$c\to +\infty$ up to an order consistent with the remainder $O(3,4)$ in
(\ref{eq:4.1}), and integrate over $z$ (using (\ref{eq:4.8}) below).
Thus we have to consider integrals involving $\overline\Lambda^{j\mu}$
and its time-derivatives, and multiplied by some {\it even} powers of
$|{\bf y}|$.  The structure of $\overline\Lambda^{j\mu}$ as given by
(\ref{eq:2.12b}) and (\ref{eq:2.12c}) is of the type $\partial V\partial
V$, where $V$ represents a retarded potential (\ref{eq:2.3a}) or
(\ref{eq:2.3b}) and $\partial$ is some time or space derivation (all
indices suppressed).  By expanding the retardation argument in the $V$'s
up to the same order $O(3,4)$, we see that the structure of
$\overline{\Lambda}^{j\mu}$ is of the type $\partial U\partial U$ or
$\partial U \partial X$, where $U$ and $X$ are instantaneous
(Poisson-type) potentials of the type $\int d^3 {\bf x}\, \sigma (x)
|{\bf y}-{\bf x}|^\alpha$ with $\alpha =\pm 1$ (see
(\ref{eq:4.14})-(\ref{eq:4.15}) below).  Thus, $\overline\Lambda^{j\mu}$
is composed of terms of the type $\int d^3{\bf x}_1 d^3{\bf x}_2
\partial_{x_1} \partial_{x_2} \sigma (x_1) \sigma (x_2) |{\bf y}-
{\bf x}_1|^{\alpha_1} |{\bf y}-{\bf x}_2|^{\alpha_2}$ where $\alpha_1
+\alpha_2 =-2$ or 0, and where the derivatives act on the source points
$x_1=({\bf x}_1,t)$ and $x_2=({\bf x}_2,t)$ (we use the fact that
$\partial_{\bf y} |{\bf y}-{\bf x}|^\alpha =-\partial_{\bf x} |{\bf y}-
{\bf x}|^\alpha$). Finally, it remains to commute the integration over
${\bf y}$ with the integrations over ${\bf x}_1$ and ${\bf x}_2$ to
arrive at a series of integrals of the type $\int d^3{\bf y}|{\bf y}|
^{B+\alpha_0}\hat y_{L'} |{\bf y}-{\bf x}_1|^{\alpha_1} |{\bf y}-{\bf x}_2|
^{\alpha_2}$, where $\alpha_0$ is an {\it even} integer and $\alpha_1
+\alpha_2 =-2$ or 0 (we have used $y_i\hat y_L \sim \hat y_{iL} +
|{\bf y}|^2 \delta_{i<i_\ell} \hat y_{L-1>}$). These
integrals have no poles by the lemma (\ref{eq:4.2}). Therefore, we have
proved that (\ref{eq:4.1}) reads in fact as
\begin{equation}
{d\over cdt} \overline{\cal T}^{0\mu}_L (t) = \ell
\,\overline{\cal T}^{\mu\langle i_\ell}_{L-1\rangle}(t) +{1\over 2\ell+3}
 \left( {d\over cdt}\right)^2 \overline{\cal T}^{j\mu}_{jL} (t) + O(3,4)\ .
 \label{eq:4.3}
\end{equation}
This relation shows that the multipole expansion in the right-hand-side
of (\ref{eq:3.39}) is divergenceless modulo small terms of order $O(7,8)$.
This is consistent with the fact that the ``$q$-part'' of the external
metric is found to be zero at order $O(7,7,7)$, see (\ref{eq:2.25}) and the
appendix~\ref{sec:apa}. Because of the uncontrolled remainder in
(\ref{eq:4.3}), we
find that the four relations (5.27) in \cite{DI91'} are modified by
small post-Newtonian terms:  ${\cal C}_L$ is modified by a term $O(7)$,
and ${\cal G}_L$, ${\cal T}_L$, ${\cal J}_L$ are modified by terms
$O(8)$.  Thus the $0i$ components (5.9b) in \cite{DI91'} involve some
uncontrolled terms $O(7)$, while the $ij$ components involve uncontrolled
terms $O(8)$ (the $00$ component not being affected).  All these terms
fall into the remainder term $O(7,7,7)$ in (\ref{eq:3.39}), showing that
the end formulas (5.33) and (5.35) of \cite{DI91'} giving the expressions
of the mass-type and current-type multipole moments in linearized
gravity can be applied to our case with the replacement of the matter
stress-energy tensor $T^{\mu\nu}$ by the total stress-energy tensor
$\overline\tau^{\mu\nu}$, and with the addition of analytic continuation
factors $|{\bf y}|^B$ and of the ``finite part'' prescription.

The matching equation (\ref{eq:3.39}) can thus be written in the form
\begin{equation}
 Gh^{\mu\nu}_{\rm can(1)} [M_L ,S_L] = Gh^{\mu\nu}_{\rm can(1)} [I_L,J_L]
 + \partial\omega^{\mu\nu} +\partial\varphi^{\mu\nu}
   +O(7,7,7)\ , \label{eq:4.4}
\end{equation}
where $\partial\omega^{\mu\nu}$ is a certain harmonic linear gauge
transformation $(\Box \omega^\mu =0)$,
and where the mass-type and current-type {\it source moments} $I_L$ and
$J_L$ are given by
\begin{mathletters}
\label{eq:4.5}
\begin{eqnarray}
 I_L(t) &=& {\rm FP}_{B=0}\int d^3 {\bf y}|{\bf y}|^B
 \int^1_{-1} dz \biggl[ \delta_\ell (z) \hat y_L \overline\Sigma -
 {4(2\ell +1)\over c^2(\ell +1)(2\ell +3)}\, \delta_{\ell +1} (z)
 \hat y_{iL}\partial_t \overline\Sigma_i\nonumber \\
 &&\qquad\qquad + {2(2\ell +1)\over c^4(\ell+1)(\ell+2)(2\ell+5)}
 \delta_{\ell+2}(z) \hat y_{ijL} \partial^2_t \overline\Sigma_{ij} \biggr]
 \left( {\bf y},t+z {|{\bf y}|\over c}\right)\ , \label{eq:4.5a}\\
  &&  \nonumber \\
 J_L(t) &=& {\rm FP}_{B=0}\int d^3 {\bf y}|{\bf y}|^B
 \int^1_{-1} dz \biggl[ \delta_\ell (z) \varepsilon_{ab<i_\ell}
  \hat y_{L-1>a}\overline\Sigma_b  \nonumber\\
 &&\qquad\qquad - {2\ell +1\over c^2(\ell+2)(2\ell+3)}
 \delta_{\ell+1}(z) \varepsilon_{ab<i_\ell}\hat y_{L-1>ac} \partial_t
   \overline\Sigma_{bc} \biggr]
 \left( {\bf y},t+z {|{\bf y}|\over c}\right)\ . \label{eq:4.5b}
\end{eqnarray}
\end{mathletters}
The mass, current and stress densities in (\ref{eq:4.5}) are defined by
\begin{mathletters}
\label{eq:4.6}
\begin{eqnarray}
 \overline\Sigma &=&{\overline\tau^{00} +\overline\tau^{ii}\over c^2}\ ,
 \label{eq:4.6a}\\
 \overline\Sigma_i &=&{\overline\tau^{0i}\over c}\ ,  \label{eq:4.6b}\\
 \overline\Sigma_{ij} &=&\overline\tau^{ij}\ ,  \label{eq:4.6c}
\end{eqnarray}
\end{mathletters}
where the stress-energy tensor $\overline\tau^{\mu\nu}$ is given by
(\ref{eq:3.37}). The weighting function $\delta_\ell (z)$ is defined
in (\ref{eq:3.13}). The matching equation (\ref{eq:4.4}) then tells us
that the gauge transformation $\partial\omega^{\mu\nu}$ necessarily
satisfies $\partial\omega^{\mu\nu} = - \partial\varphi^{\mu\nu} +O(7,8)$
(indeed remember that $\Box \varphi^\mu =O(7,8)$), {\it and} that the
``canonical'' moments $M_L$, $S_L$ in the left-hand-side of (\ref{eq:4.4})
are related to the source moments $I_L$, $J_L$ by
\begin{mathletters}
\label{eq:4.7}
\begin{eqnarray}
 M_L(t) &=&I_L (t) + O(5) \ , \label{eq:4.7a}\\
 S_L(t) &=&J_L (t) + O(4) \ . \label{eq:4.7b}
\end{eqnarray}
\end{mathletters}
[The relation (\ref{eq:4.7a}) for the mass moment comes from the 00
component of the matching equation (\ref{eq:4.4}), while the relation
(\ref{eq:4.7b}) for the current moment comes from the $0i$ components
of (\ref{eq:4.4}).] The relations (\ref{eq:4.7a}) and (\ref{eq:4.7b})
are exactly the ones needed to solve the wave generation problem at the
second-post-Newtonian approximation.

\subsection{Expressions of the mass-type and current-type source moments}
\label{ssec:4.2}

Since the relations (\ref{eq:4.7}) are only valid up to some post-Newtonian
order, it is sufficient to consider the post-Newtonian expansion
$c\to +\infty$ of the source moments $I_L$, $J_L$ defined by (\ref{eq:4.5}).
This expansion is achieved by means of a formula derived in the
appendix~B of \cite{BD89}, and giving the expansion when $c\to +\infty$
of terms involving the average over $z$ appearing in (\ref{eq:4.5}).
This formula reads as
\begin{eqnarray}
 \int^1_{-1} dz \delta_\ell (z) \overline\Sigma \left( {\bf y},t+z
   {|{\bf y}|\over c} \right) &=&\overline\Sigma ({\bf y},t) +
 {|{\bf y}|^2\over 2c^2 (2\ell+3)} \partial^2_t \overline\Sigma ({\bf y},t)
 \nonumber \\
  &&\qquad + {|{\bf y}|^4\over 8c^4 (2\ell+3)(2\ell+5)}
    \partial^4_t \overline\Sigma ({\bf y},t) + O(6)\ . \label{eq:4.8}
\end{eqnarray}
Using (\ref{eq:4.8}), and retaining consistently the powers of $c^{-1}$,
we obtain
\begin{eqnarray}
 I_L(t) &=& {\rm FP}_{B=0}\int d^3 {\bf y}|{\bf y}|^B
  \biggl[ \hat y_L \overline\Sigma +
 {|{\bf y}|^2\hat y_{L} \over 2c^2(2\ell +3)}\,
 \partial^2_t \overline\Sigma\nonumber \\
 && +{|{\bf y}|^4\hat y_L\over 8c^4(2\ell+3)(2\ell+5)} \partial^4_t
  \overline\Sigma - {4(2\ell+1) \hat y_{iL}\over
  c^2(\ell+1)(2\ell+3)} \partial_t\overline\Sigma_i \nonumber\\
  && -{2(2\ell+1)|{\bf y}|^2\hat y_{iL}\over c^4(\ell+1)(2\ell+3)(2\ell+5)}
   \partial^3_t\overline\Sigma_i
  + {2(2\ell+1)\hat y_{ijL}\over c^4(\ell+1)(\ell+2)(2\ell+5)}
   \partial^2_t\overline\Sigma_{ij} \biggr] ({\bf y},t) +O(6)\ ,
\label{eq:4.9}\\
  && \nonumber \\
 J_L(t) &=& {\rm FP}_{B=0}\,\varepsilon_{ab<i_\ell}
   \int d^3 {\bf y}|{\bf y}|^B
   \biggl[ \hat y_{L-1>a}\overline\Sigma_b + {|{\bf y}|^2\hat y_{L-1>a}\over
   2c^2 (2\ell+3)} \partial^2_t \overline\Sigma_b \nonumber \\
 &&\qquad\qquad - {(2\ell +1)\hat y_{L-1>ac}\over c^2(\ell+2)(2\ell+3)}
 \partial_{t}\overline\Sigma_{bc} \biggr] ({\bf y},t) + O(4)\ . \label{eq:4.10}
\end{eqnarray}
We must then insert into (\ref{eq:4.9})-(\ref{eq:4.10}) explicit formulas
of $\overline\Sigma$, $\overline\Sigma_i$ and $\overline\Sigma_{ij}$
which are easily computed from (\ref{eq:3.37}) where we use the
expression (\ref{eq:2.11}) of $\overline\lambda =|g| +O(6)$ and the
expressions (\ref{eq:2.12}) of the components of the effective nonlinear
source $\overline\Lambda^{\mu\nu}$. We find the formulas
\begin{mathletters}
\label{eq:4.11}
\begin{eqnarray}
 \overline\Sigma &=& \left[ 1+{4V\over c^2} -{8\over c^4} (W-V^2)\right]
 \sigma - {1\over \pi Gc^2}\,\partial_i V \partial_i V \nonumber \\
 && + {1\over \pi Gc^4} \biggl\{ -V \partial^2_t V- 2V_i\partial_t
   \partial_i V - W_{ij}\partial^2_{ij} V-{1\over 2} (\partial_t V)^2
    \nonumber \\
 && \qquad +2 \partial_i V_j \partial_j V_i + 2\partial_i V\partial_i W
  - {7\over 2} V\partial_i V\partial_i V \biggr\}\ , \label{eq:4.11a}\\
 \overline\Sigma_i &=& \left[ 1 +{4V\over c^2}\right] \sigma_i  +
  {1\over \pi Gc^2} \left\{ \partial_k V(\partial_i V_k -\partial_k V_i)
  + {3\over 4} \partial_t V \partial_i V \right\} + O(4)\ , \label{eq:4.11b}\\
 \overline\Sigma_{ij} &=& \sigma_{ij} + {1\over 4\pi G} \left\{ \partial_i
 V \partial_j V - {1\over 2} \delta_{ij} \partial_k V \partial_k V\right\}
  + O(2)\ . \label{eq:4.11c}
\end{eqnarray}
\end{mathletters}
Again retaining the powers of $c^{-1}$ consistently with the accuracy
indicated in (\ref{eq:4.9})-(\ref{eq:4.10}), we can now write
formulas for the source moments in raw form:
\begin{eqnarray}
 I_L(t) &=& {\rm FP}_{B=0}\int d^3 {\bf y}|{\bf y}|^B
 \biggl\{ \biggl[1 + {4\over c^2}V-{8\over c^4} (W-V^2)\biggr] \hat y_L
 \sigma - {1\over \pi Gc^2} \hat y_L \partial_i V \partial_i V\nonumber \\
 && + {1\over \pi Gc^4}\hat y_L \biggl[ -V \partial^2_t V- 2V_i\partial_t
   \partial_i V - W_{ij}\partial^2_{ij} V-{1\over 2} (\partial_t V)^2
    \nonumber \\
 &&\qquad\quad\qquad +2 \partial_iV_j\partial_j V_i+2\partial_iV\partial_i W
  - {7\over 2} V\partial_i V\partial_i V \biggr]\  \nonumber\\
 && +{|{\bf y}|^2\hat y_L\over 2c^2(2\ell+3)} \partial^2_t
   \left[ \left( 1+{4V\over c^2}\right) \sigma - {1\over \pi Gc^2}
   \partial_i V\partial_i V \right] \nonumber \\
 && +{|{\bf y}|^4\hat y_L\over 8c^4(2\ell+3)(2\ell+5)} \partial^4_t\sigma
   - {2(2\ell+1)|{\bf y}|^2\hat y_{iL}\over c^4(\ell+1)(2\ell+3)(2\ell+5)}
   \partial^3_t\sigma_i  \nonumber\\
 && -{4(2\ell+1)\hat y_{iL}\over c^2(\ell+1)(2\ell+3)}    \partial_t
   \left[ \left( 1 +{4V\over c^2} \right) \sigma_i  +
  {1\over \pi Gc^2} \left\{ \partial_k V(\partial_i V_k -\partial_k V_i)
  + {3\over 4} \partial_t V \partial_i V \right\} \right] \nonumber \\
 && + {2(2\ell+1)\hat y_{ijL}\over c^4(\ell+1)(\ell+2)(2\ell+5)}
    \partial_t^2 \left[ \sigma_{ij} + {1\over 4\pi G} \partial_i V
    \partial_j V \right] \biggr\} + O(6)\ , \label{eq:4.12} \\
 && \nonumber \\
 J_L(t) &=& {\rm FP}_{B=0}\,\varepsilon_{ab<i_\ell}
   \int d^3 {\bf y}|{\bf y}|^B
   \biggl\{ \hat y_{L-1>a}\left( 1+{4\over c^2}V\right) \sigma_b
  + {|{\bf y}|^2\hat y_{L-1>a}\over
   2c^2 (2\ell+3)} \partial^2_t \sigma_b \nonumber \\
 && \qquad\qquad + {1\over \pi Gc^2} \hat y_{L-1>a}
  \left[ \partial_k V(\partial_b V_k -\partial_k V_b)
  + {3\over 4} \partial_t V \partial_b V \right] \nonumber \\
 &&\qquad\qquad - {(2\ell +1)\hat y_{L-1>ac}\over c^2(\ell+2)(2\ell+3)}
 \partial_{t}\left[ \sigma_{bc} +{1\over 4\pi G}  \partial_b V\partial_c
   V \right] \biggr\} + O(4)\ .  \label{eq:4.13}
\end{eqnarray}
Recall that the analytic continuation procedure in
(\ref{eq:4.12})-(\ref{eq:4.13}) is needed only for handling the (apparently)
non-compact supported terms like the term $\hat y_L\partial_i V\partial_i
V$ in (\ref{eq:4.12}), and could be removed from manifestly compact
supported terms like e.g. $\hat y_L \sigma V^2$.

Let us now derive an equivalent but somewhat simpler form for the mass-type
multipole source moment (\ref{eq:4.12}). First we replace the retarded
potentials $V$, $V_i$, $W_{ij}$ used in (\ref{eq:4.12}) by
``instantaneous'' potentials $U$, $X$, $U_i$, $P_{ij}$ (and $P=P_{ii}$)
defined by the post-Newtonian expansions
\begin{mathletters}
\label{eq:4.14}
\begin{eqnarray}
 V &=& U + {1\over 2c^2}\, \partial^2_t X + O(3)\ , \label{eq:4.14a} \\
 V_i &=& U_i + O(2)\ , \label{eq:4.14b} \\
 W_{ij} &=& P_{ij} + O(1)\ . \label{eq:4.14c}
\end{eqnarray}
\end{mathletters}
These instantaneous potentials are
\begin{mathletters}
\label{eq:4.15}
\begin{eqnarray}
 U({\bf x},t) &=& G \int {d^3{\bf x}'\over |{\bf x}-{\bf x}'|} \sigma
 ({\bf x}',t)\ , \label{eq:4.15a}\\
 X({\bf x},t) &=& G \int d^3{\bf x}'|{\bf x}-{\bf x}'| \sigma
 ({\bf x}',t)\ , \label{eq:4.15b}\\
 U_i({\bf x},t) &=& G \int {d^3{\bf x}'\over |{\bf x}-{\bf x}'|} \sigma_i
 ({\bf x}',t)\ , \label{eq:4.15c}\\
 P_{ij}({\bf x},t) &=& G \int {d^3{\bf x}'\over |{\bf x}-{\bf x}'|}
  \left[ \sigma_{ij} + {1\over 4\pi G} \left( \partial_i U\partial_j U
 - {1\over 2} \delta_{ij} \partial_k U\partial_k U\right) \right]
  ({\bf x}',t)\ , \label{eq:4.15d}\\
P({\bf x},t) &=& G \int {d^3{\bf x}'\over |{\bf x}-{\bf x}'|}
  \left[ \sigma_{ii} - {1\over 2} \sigma U \right]
  ({\bf x}',t)  +  {{U^2} \over {4}} \ . \label{eq:4.15e}
\end{eqnarray}
\end{mathletters}
We can then rewrite (\ref{eq:4.12}) as
\begin{eqnarray}
 I_L(t) &=& {\rm FP}_{B=0}\int d^3 {\bf y}|{\bf y}|^B
 \biggl\{ \biggl[1 + {4\over c^2}U+{1\over c^4} (2\partial^2_t X -8P +8U^2)
 \biggr] \hat y_L  \sigma  \nonumber \\
 && - {1\over \pi Gc^2} \hat y_L \partial_i U \partial_i U
 + {1\over \pi Gc^4}\hat y_L \biggl[ - \partial_i U\partial_i \partial^2_tX
 - U\partial^2_tU \nonumber\\
 && -2U_i\partial_t \partial_i U - P_{ij}\partial^2_{ij} U-{1\over 2}
   (\partial_t U)^2 + 2\partial_i U_j \partial_j U_i + 2\partial_i U
   \partial_iP
  - {7\over 2} U\partial_i U\partial_i U \biggr]\  \nonumber\\
 && +{|{\bf y}|^2\hat y_L\over 2c^2(2\ell+3)} \partial^2_t
   \left[ \left( 1+{4U\over c^2}\right) \sigma - {1\over \pi Gc^2}
   \partial_i U\partial_i U \right] \nonumber \\
 && +{|{\bf y}|^4\hat y_L\over 8c^4(2\ell+3)(2\ell+5)} \partial^4_t\sigma
   - {2(2\ell+1)|{\bf y}|^2\hat y_{iL}\over c^4(\ell+1)(2\ell+3)(2\ell+5)}
   \partial^3_t\sigma_i  \nonumber\\
 && -{4(2\ell+1)\hat y_{iL}\over c^2(\ell+1)(2\ell+3)}    \partial_t
   \left[ \left( 1 +{4U\over c^2} \right) \sigma_i  +
  {1\over \pi Gc^2} \left\{ \partial_k U(\partial_i U_k -\partial_k U_i)
  + {3\over 4} \partial_t U \partial_i U \right\} \right] \nonumber \\
 && + {2(2\ell+1)\hat y_{ijL}\over c^4(\ell+1)(\ell+2)(2\ell+5)}
  \partial_t^2 \left[ \sigma_{ij} + {1\over 4\pi G} \partial_i U
  \partial_j U \right] \biggr\} + O(5)\ .  \label{eq:4.16}
\end{eqnarray}

Several reductions of this expression can be done by integrating various
terms by parts. For instance we can operate by part the term $\sim \hat
y_L \partial_i U\partial_i U$ using $\bigtriangleup \hat y_L=0$ and
$\bigtriangleup U=-4\pi G\sigma$.
 This yields
\begin{equation}
  {\rm FP}_{B=0}\int d^3 {\bf y}|{\bf y}|^B
  \hat y_L \partial_i U\partial_i U =
  {\rm FP}_{B=0}\int d^3 {\bf y}|{\bf y}|^B
  \left\{ 4\pi G \hat y_L \sigma U + {1\over 2} \partial_i
 \left[ \hat y_L \partial_i U^2 - \partial_i \hat y_L U^2 \right]
  \right\}\ . \label{eq:4.17}
\end{equation}
The second term in the right-hand-side of (\ref{eq:4.17}), which is
made of the product of $|{\bf y}|^B$ and of a pure divergence, is easily
shown to be zero thanks to the lemma (\ref{eq:4.2}). Indeed,
the differentiation of the factor $|{\bf y}|^B$ yields $\alpha_0=-2$, and
since the term involves two potentials $U$ it has $\alpha_1 +\alpha_2
=-2$. Hence we can write
\begin{equation}
  {\rm FP}_{B=0}\int d^3 {\bf y}|{\bf y}|^B
  \hat y_L \partial_i U\partial_i U = 4\pi G \int d^3 {\bf y} \hat y_L
  \sigma U\ , \label{eq:4.18}
\end{equation}
where we have removed the factor $|{\bf y}|^B$ in the right-hand-side
since the term is compact-supported. Note that the use of the lemma
(\ref{eq:4.2}) permits to freely integrate by parts all non-compact-supported
terms in (\ref{eq:4.16}) except the term involving the product
of {\it three} $U$'s, i.e. the term $\sim \hat y_L U\partial_i U\partial_i U$,
which must be treated separately. For this term we write, like in
(\ref{eq:4.17}),
\begin{equation}
  {\rm FP}_{B=0}\int d^3 {\bf y}|{\bf y}|^B
  \hat y_L U \partial_i U\partial_i U =
  {\rm FP}_{B=0}\int d^3 {\bf y}|{\bf y}|^B
  \left\{ 2\pi G \hat y_L \sigma U^2 + {1\over 6} \partial_i
 \left[ \hat y_L \partial_i U^3 - \partial_i \hat y_L U^3 \right]
  \right\}\ . \label{eq:4.19}
\end{equation}
To cancel the second term in the right-hand-side, we have to show that
the integral $\int d^3 {\bf y} |{\bf y}|^{B-2} \hat y_{L'} |{\bf y}-{\bf
x}_1|^{-1}$ $|{\bf y}-{\bf x}_2|^{-1}$ $|{\bf y}-{\bf x}_3|^{-1}$
has no pole at $B=0$ (this case is not covered by (\ref{eq:4.2})).
When $|{\bf y}|\to +\infty$ the integrand behaves like $|{\bf y}|^{B
+\ell'-\ell_1 -\ell_2 -\ell_3 -5} \hat n_{L'} \hat n_{L_1} \hat n_{L_2}
\hat n_{L_3}$ (where $n_i = y_i/|{\bf y}|$). The angular integration
shows that $\ell_1+\ell_2+\ell_3 =\ell'+2p$ where $p$ is a {\it positive}
integer, so that the remaining radial integral is $\int d|{\bf y}||{\bf y}|
^{B-2p-3}$ which cannot have a pole (this would mean $p=-1$). Thus we
can conclude, like in (\ref{eq:4.18}), that
\begin{equation}
 {\rm FP}_{B=0}\int d^3 {\bf y}|{\bf y}|^B
  \hat y_L U \partial_i U\partial_i U = 2\pi G \int d^3 {\bf y} \hat y_L
  \sigma U^2 \ . \label{eq:4.20}
\end{equation}
Incidentally, the above proof shows that in higher post-Newtonian
approximations, terms can arise which cannot be integrated by parts
without generating poles. This is for instance the case of the term
$\sim \hat y_L U\partial_i U\partial_i \partial^2_t X$ which is expected
to arise at the $O(6)$ level.

 The identity (\ref{eq:4.18}) shows that the term $\sim \hat y_L
\partial_i U\partial_i U$ in (\ref{eq:4.16}) exactly cancels a previous
term $\sim \hat y_L \sigma U$, so that we recover, at the first
post-Newtonian approximation, the expression obtained in \cite{BD89},
i.e. (2.27) in \cite{BD89}. We use
also the identity (\ref{eq:4.20}), and then perform several other
manipulations like for instance one showing that the two terms involving the
potential
$X$ in (\ref{eq:4.16}) can be advantageously replaced by a single term
$\sim \hat y_L U\partial^2_t U$ ---~all these manipulations are
justified by the lemma (\ref{eq:4.2}). They yield the
expression
\begin{eqnarray}
 I_L(t) &=& {\rm FP}_{B=0}\int d^3 {\bf y}|{\bf y}|^B
 \biggl\{ \hat y_L \biggl[\sigma + {4\over c^4}(\sigma_{ii}U -\sigma P)
 \biggr] + {|{\bf y}|^2\hat y_L\over 2c^2(2\ell+3)} \partial^2_t
\sigma\nonumber\\
 && -{4(2\ell+1)\hat y_{iL}\over c^2(\ell+1)(2\ell+3)} \partial_t
   \left[ \left( 1 +{4U\over c^2} \right) \sigma_i  +
  {1\over \pi Gc^2} \left( \partial_k U[\partial_i U_k -\partial_k U_i]
  + {3\over 4} \partial_t U \partial_i U \right) \right] \nonumber \\
 && +{|{\bf y}|^4\hat y_L\over 8c^4(2\ell+3)(2\ell+5)} \partial^4_t\sigma
   - {2(2\ell+1)|{\bf y}|^2\hat y_{iL}\over c^4(\ell+1)(2\ell+3)(2\ell+5)}
   \partial^3_t\sigma_i  \nonumber\\
 && + {2(2\ell+1)\over c^4(\ell+1)(\ell+2)(2\ell+5)} \hat y_{ijL}
   \partial_t^2 \left[ \sigma_{ij} + {1\over 4\pi G} \partial_i U
   \partial_j U \right] \nonumber\\
 &&  + {1\over \pi Gc^4}\hat y_L \biggl[ -P_{ij} \partial_{ij}^2 U
  - 2 U_i\partial_t\partial_i U + 2\partial_i U_j \partial_j U_i
  - {3\over 2} (\partial_t U)^2 -U\partial_t^2 U\biggr] \biggr\}
   + O(5)\ . \label{eq:4.21}
\end{eqnarray}
The expression (\ref{eq:4.21}) will be applied, in a forthcoming paper
\cite{BDIWW94}, to the problem of the generation of waves by a coalescing
compact binary at the 2-PN approximation. Note that many other
transformations of the expression (\ref{eq:4.21}) could be done using the
equations of motion and of conservation of mass. In the
appendix~\ref{sec:apb} we show that in the case $\ell =0$ the expression
(\ref{eq:4.21}) reduces to the expression of the conserved total mass
$M$ at 2-PN which is known from the equation of conservation
of mass at 2-PN.

Finally we prove that the expression (\ref{eq:4.13}) of the current-type
source moment is equivalent to the expression obtained in \cite{DI91}.
We make the comparison with the expression (5.18) in
\cite{DI91}, which involves a STF tensor $Y^L({\bf y}_1,{\bf y}_2)$
depending on two source points ${\bf y}_1$, ${\bf y}_2$ and defined by
\begin{mathletters}
\label{eq:4.22}
\begin{equation}
 Y^L ({\bf y}_1, {\bf y}_2) = |{\bf y}_1 - {\bf y}_2| \int^1_0 d\alpha
 y_\alpha^{\langle L\rangle} \ , \label{eq:4.22a}
\end{equation}
where $y^i_\alpha = \alpha y^i_1 +(1-\alpha) y^i_2$, and where
$y^{\langle L\rangle}_\alpha$ denotes the STF part of $y^L_\alpha =
y^{i_1}_\alpha\cdots y^{i_\ell}_\alpha$. An alternative form of $Y^L$ can
be obtained by explicitly performing the integration over $\alpha$. It reads
\begin{equation}
 Y^L ({\bf y}_1, {\bf y}_2) = {|{\bf y}_1 - {\bf y}_2|\over \ell+1}
   \sum^\ell_{p=0}
 y_1^{\langle L-P} y_2^{P\rangle} \ , \label{eq:4.22b}
\end{equation}
\end{mathletters}
where we sum over the number $p$ of indices present on $y_2^P =y^{i_1}_2
\cdots y^{i_p}_2$ (in which case $y^{L-P}_1 = y^{i_{p+1}}_1 \cdots
y^{i_\ell}_1$) and without $p$-dependent coefficient in the sum, e.g.,
$Y^{ij} = {1\over 3} |{\bf y}_1-{\bf y}_2|$ $\times (y_1^{\langle
i}y_1^{j\rangle}
+y_1^{\langle i} y_2^{j\rangle} + y_2^{\langle i} y_2^{j\rangle})$. Then
the formula which permits to relate our work with the formalism used in
\cite{DI91} is
\begin{equation}
 {\rm FP}_{B=0} \int {d^3{\bf y} |{\bf y}|^B \hat y_L
 \over |{\bf y}-{\bf y}_1| |{\bf y}-{\bf y}_2|} = - 2\pi Y^L ({\bf y}_1,
  {\bf y}_2)\ . \label{eq:4.23}
\end{equation}
The proof of this formula is as follows. We know from (3.9c) in
\cite{DI91} that $(|{\bf y}-{\bf y}_1||{\bf y}-{\bf y}_2|)^{-1} =
\Delta_{\bf y} k-2\pi \delta_{12}$ where $\delta_{12} =|{\bf y}_1
-{\bf y}_2| \int^1_0 d\alpha \delta({\bf y} -{\bf y}_\alpha)$ represents
a Dirac distribution on the segment joining ${\bf y}_1$ and ${\bf y}_2$
($\delta$ is the usual three-dimensional Dirac distribution),
and where $k$ is some kernel given by $k={1\over 2} \ln [(|{\bf y}
-{\bf y}_1| + |{\bf y}-{\bf y}_2|)^2-|{\bf y}_1-{\bf y}_2|^2].$ By
substitution into the left-hand-side of (\ref{eq:4.23}) one gets two terms.
The first one is the finite part at $B=0$ of the integral $\int d^3
{\bf y}|{\bf y}|^B\hat y_L \Delta_{\bf y} k$, which is also equal to
$B\int d^3{\bf y}|{\bf y}|^{B-2} y_i [\partial_i \hat y_L k - \hat y_L
\partial_i k]$. We replace into the latter integral the convergent Taylor
expansion
when ${\bf y}_1$, ${\bf y}_2 \to 0$ of the kernel $k$ (which is
analytic in ${\bf y}_1$, ${\bf y}_2$; see (3.14) in \cite{DI91}), and
find that the only remaining radial integrals are of the type $B \int
d^3{\bf y}|{\bf y}|^{B-2k}$ or $B\int d^3{\bf y}|{\bf y}|^{B-2} \ln
|{\bf y}|$, where $k$ is an integer. These radial integrals are zero
at B=0 (no poles). Thus it remains only the second term which is the
finite part at $B=0$ of $\int d^3 {\bf y}|{\bf y}|^B \hat y_L(-2\pi
\delta_{12})$, and readily yields the result (\ref{eq:4.23}). By
multiplying (\ref{eq:4.23}) by some densities $\sigma ({\bf y}_1,t)$ and
$\sigma({\bf y}_2,t)$ and integrating over $d^3{\bf y}_1$ and $d^3{\bf y}_2$
we obtain such relations as
\begin{equation}
 {\rm FP}_{B=0} \int d^3{\bf y} |{\bf y}|^B \hat y_L
 \partial_iU\partial_jU = -2 \pi \int\!\!\!\int d^3{\bf y}_1d^3{\bf y}_2
 \sigma (y_1) \sigma (y_2) Y^L_{,ij}
 ({\bf y}_1,{\bf y}_2)\ , \label{eq:4.24}
\end{equation}
where $Y^L_{,ij} =\partial^2 Y^L/\partial y^i_1 \partial y^j_2$, showing the
complete equivalence between our expression (\ref{eq:4.13}) above (where
the $V$'s can be replaced by the corresponding $U$'s) and
the expression (5.18) in \cite{DI91}.

\subsection{The asymptotic waveform at the 2-PN approximation}
\label{sec:4.3}

It has been shown in \cite{B87} (see also \cite{B87'}) that the
``canonical" external field (\ref{eq:2.16}), which satisfies all-over
$D_e$ the harmonic-coordinates Einstein's equations
(\ref{eq:1.1})-(\ref{eq:1.2}) (in the sense of formal nonlinear
expansions), can be rewritten in a so-called {\it radiative} coordinate
system $X^\mu =(cT,{\bf X})$ in which it is of the Bondi type at large
distances from the source. It is sufficient to consider the
transverse-traceless (TT) projection of the leading-order $1/R$ part of
the spatial metric (where $R=|{\bf X}|$ is the distance to the source).
Denoting by $h^{TT}_{km} (X^\mu) = (g_{km}(X^\mu) -\delta_{km})^{TT}$ this
TT projection of the spatial metric (where $g_{km}$ is the usual covariant
metric), we can then {\it uniquely} decompose the $1/R$ part of
$h^{TT}_{km}$ into the infinite multipole moment series \cite{Th80}
\begin{eqnarray}
 h^{TT}_{km} ({\bf X},T) =&& {4G\over c^2R} {\cal P}_{ijkm} ({\bf N})
 \sum^\infty_{\ell =2} {1\over c^\ell \ell !} \biggl\{ N_{L-2} U_{ijL-2}\!
 \left( T-R/ c\right) \nonumber \\
 &&- {2\ell \over (\ell +1)c} N_{aL-2}
 \varepsilon_{ab(i} V_{j)bL-2}\!\left( T-R/ c\right)\biggr\} +
 O \left( {1\over R^2}\right)\ , \label{eq:4.25}
\end{eqnarray}
where the {\it radiative} moments $U_L$ and $V_L$ represent two infinite
sets of functions of the retarded time $T-R/c$, which are STF in their
indices $L=i_1\cdots i_\ell$ ($\ell$ goes from 2 up to infinity). These
functions {\it by definition} parametrize the asymptotic waveform. The
coefficients in (\ref{eq:4.25}) have been chosen so that the moments
$U_L$ and $V_L$ reduce, in the non-relativistic limit $c\to +\infty$
to the $\ell$th time-derivatives of the usual ``Newtonian" mass-type
and current-type moments of the source \cite{Th80}. Our notation in
(\ref{eq:4.25}) is $N_{L-2}= N_{i_1}\cdots N_{i_{\ell -2}}$ with $N_i
=X^i/R$, $N_{aL-2} =N_a N_{L-2}$, $T_{(ij)}={1\over 2} (T_{ij} +T_{ji})$,
and
\begin{equation}
 {\cal P}_{ijkm} ({\bf N}) = (\delta_{ik} -N_iN_k)(\delta_{jm} -N_jN_m)
 -{1\over 2} ( \delta_{ij} -N_iN_j) (\delta_{km}-N_kN_m)\ .
 \label{eq:4.26}
\end{equation}

 At the 2-PN approximation, including all terms up to the level
$\varepsilon^4 \sim c^{-4}$, the waveform (\ref{eq:4.25}) reads as
\begin{eqnarray}
 h^{TT}_{km} = {2G\over c^4R} {\cal P}_{ijkm} \biggl\{  U_{ij}
  &+& {1\over c} \left[ {1\over 3} N_a U_{ija} + {4\over 3}
   \varepsilon_{ab(i} V_{j)a} N_b \right] \nonumber \\
  &+& {1\over c^2} \left[ {1\over 12} N_{ab} U_{ijab} + {1\over 2}
   \varepsilon_{ab(i} V_{j)ac} N_{bc} \right] \nonumber \\
  &+& {1\over c^3} \left[ {1\over 60} N_{abc} U_{ijabc} + {2\over 15}
   \varepsilon_{ab(i} V_{j)acd} N_{bcd} \right] \nonumber \\
  &+& {1\over c^4} \left[ {1\over 360} N_{abcd} U_{ijabcd} + {1\over 36}
   \varepsilon_{ab(i} V_{j)acde} N_{bcde} \right] + O(5) \biggr\}
   \ .\label{eq:4.27}
\end{eqnarray}
By differentiating, squaring and averaging over angles this expression,
one obtains the energy loss formula at the 2-PN approximation, giving
the rate of decrease of the Bondi energy $E_B$:
\begin{eqnarray}
 {dE_B\over dT} = - {G\over c^5} \biggl\{ {1\over 5} U^{(1)}_{ij} U^{(1)}_{ij}
 &+&{1\over c^2} \left[ {1\over 189} U^{(1)}_{ijk} U^{(1)}_{ijk}
 + {16\over 45} V^{(1)}_{ij} V^{(1)}_{ij}\right] \nonumber \\
 &+&{1\over c^4} \left[ {1\over 9072} U^{(1)}_{ijkm} U^{(1)}_{ijkm}
 + {1\over 84} V^{(1)}_{ijk} V^{(1)}_{ijk}\right] + O(6) \biggr\}\ .
 \label{eq:4.28}
\end{eqnarray}

 Now the point is that the external field (\ref{eq:2.16}) is
algorithmically constructed in \cite{BD86} from the linearized metric
(\ref{eq:2.17})-(\ref{eq:2.18}), which is parametrized by the ``canonical"
multipole moments $M_L$, $S_L$, and that the coordinate transformation
between the harmonic coordinates and the radiative ones can also be
algorithmically implemented \cite{B87,B87'}. Therefore, the radiative
moments $U_L$ and $V_L$ parametrizing the multipole expansion of the
asymptotic waveform (\ref{eq:4.25}) are necessarily given as some
algorithmically computable functionals of the canonical moments $M_L$
and $S_L$. It has been shown in previous papers \cite{BD89,BD92} that
$U_L$ and $V_L$ are given by some nonlinear infinite expansions in $G$
(consistently with our whole approach) of the type
\begin{mathletters}
\label{eq:4.29}
\begin{eqnarray}
 U_L (T) &=& M_L^{(\ell)} (T) + \sum_{n\geq 2} {G^{n-1}\over
  c^{3(n-1)+\Sigma \underline{\ell}_i-\ell}} X_{nL} (T)\ ,\label{eq:4.29a}\\
\varepsilon_{ai_\ell i_{\ell -1}} V_{aL-2} (T) &=&
  \varepsilon_{ai_\ell i_{\ell -1}} S^{(\ell -1)}_{aL-2} (T)
   + \sum_{n\geq 2} {G^{n-1}\over
c^{3(n-1)+\Sigma \underline{\ell}_i-\ell}} Y_{nL} (T)\ , \label{eq:4.29b}
\end{eqnarray}
\end{mathletters}
where $M^{(\ell)}_L(T)$ and $S^{(\ell-1)}_{L-1}(T)$ denote the $\ell$th
and $(\ell -1)$th time-derivatives of $M_L$ and $S_{L-1}$ computed at the
radiative time $T$, and where $X_{nL} (T)$ and $Y_{nL}(T)$ are nonlinear
functionals of order $n$ of the moments $M_L$ and $S_L$ and their
time-derivatives. The general structure of $X_{nL}$ and $Y_{nL}$ is
\begin{equation}
X_{nL} (T), Y_{nL} (T) = \sum \int^T_{-\infty} dU_1 \cdots
\int^T_{-\infty} dU_n {\cal K}_{L\underline{L}_1 \cdots \underline{L}_n}
(T,U_1,\cdots, U_n) {\cal P}^{(a_1)}_{\underline{L}_1} (U_1) \cdots
{\cal P}^{(a_n)}_{\underline{L}_n}(U_n) \ , \label{eq:4.30}
\end{equation}
where ${\cal P}^{(a)}_{\underline{L}}$ denotes the $a$th time-derivative of
either a mass moment $M_L$ (in which case $\underline{\ell}=\ell$)
or a current moment $\varepsilon_{ai_{\ell+1} i_\ell} S_{aL-1}$
endowed with a Levi-Civita symbol (in which case $\underline{\ell}=\ell+1$).
The tensor ${\cal K}_{L\underline{L}_1 \cdots \underline{L}_n}$ denotes
some dimensionless kernel whose index structure is made out only of
Kronecker symbols, and which depends only on variables having the
dimension of time.  The powers of $G$ and $c$ in (\ref{eq:4.29}) are
obtained by a simple dimensional argument, namely that the mass and current
moments $M_L$ and $S_L$ have the usual dimensions of multipole moments.
The notation $\Sigma \underline{\ell}_i$ is for $\Sigma^n_{i=1}
\underline{\ell}_i = \Sigma^n_{i=1} \ell_i +s$, where $\Sigma^n_{i=1}
\ell_i$ is the total number of indices present on the $n$ moments ${\cal
P}_{\underline{L}}$ in (\ref{eq:4.30}), and where $s$ is the number of
current moments among these $n$ moments.  As the tensor ${\cal
K}_{L\underline{L}_1\cdots \underline{L}_n}$ represents an operation of
complete contraction between the indices $L, \underline{L}_1,\cdots,
\underline{L}_n$, we have necessarily the equality
\begin{equation}
 \sum^n_{i=1} \underline{\ell}_i = \ell + 2k \ , \label{eq:4.31}
\end{equation}
where $k$ is the number of contractions among the indices
$\underline{L}_1, \cdots \underline{L}_n$.

In view of the explicit powers of $c^{-1}$ in front of the multipole moment
contributions present in (\ref{eq:4.27}), we need to compute the
relations (\ref{eq:4.29}) linking the radiative and canonical moments
only up to some definite order in $c^{-1}$. Namely, $U_{ij}$ is to be
computed up to $c^{-4}$ inclusively, $U_{ijk}$ and $V_{ij}$ are to be
computed up to $c^{-3}$, $U_{ijkm}$ and $V_{ijk}$ up to $c^{-2}$, and so
on. Now the equality (\ref{eq:4.31}) shows that the nonlinear terms in
(\ref{eq:4.29}) having $n\geq 3$, and thus coming from the cubic and
higher nonlinearities of Einstein's equations, are at least of order
$O(c^{-6}) = O(6)$ and can be neglected for our purpose. Furthermore,
the terms with $n=2$ must have $k=0$ in (\ref{eq:4.31}) since for
$k\geq 1$ the corresponding order is $O(5)$, also negligible for our
purpose. Finally, the remaining nonlinear terms with $n=2$ and $k=0$,
which represent corrections $\sim c^{-3}$ in the radiative moments, are
to be computed only in the quadrupole and octupole mass moments $U_{ij}$
and $U_{ijk}$, and in the quadrupole current moment $V_{ij}$.  We then
easily arrive at the only possibilities $\ell =\underline{\ell}_1
+\underline{\ell}_2$ (see (\ref{eq:4.31})) with $\ell =2$ (case of
$U_{ij}$) and $\underline{\ell}_1=0$, $\underline{\ell}_2=2$, or $\ell
=3$ (cases of $U_{ijk}$ and $V_{ij}$) and $\underline{\ell}_1 =0$,
$\underline{\ell}_2 =3$ or $\underline{\ell}_1 =1$, $\underline{\ell}_2
=2$.  (Indeed, one of the moments in (\ref{eq:4.30}) is necessarily
non-static, $\underline{\ell}_2 \geq 2$ say.) This corresponds to the
interaction of the mass monopole $M$ and of the mass quadrupole $M_{ij}$
(case of $U_{ij}$), of $M$ and of the mass octupole $M_{ijk}$ or of the
mass dipole $M_i$ and of $M_{ij}$ (case of $U_{ijk}$), and of $M$ and
the current quadrupole $S_{ij}$ (case of $V_{ij}$).  Let us combine
these informations with the results of \cite{BD92} showing that two (and
only two) types of ``hereditary'' contributions arise in the radiative
moments $U_L$, $V_L$ at the quadratic nonlinear approximation, namely
the ``tail" contributions involving the interaction between $M$ and
non-static moments $M_L$ or $S_L$, and the ``memory" contribution
involving the interaction between two non-static moments $M_L$.  By the
previous reasoning, the latter ``memory" contribution can be neglected,
and the former ``tail'' contributions need to be included only in the
radiative moments $U_{ij}$, $U_{ijk}$ and $V_{ij}$.  Hence we can write,
from (2.42) and (3.4) in \cite{BD92},
\begin{mathletters}
\label{eq:4.32}
\begin{eqnarray}
 U_{ij} (T) &=& M^{(2)}_{ij} (T) + {2GM\over c^3} \int^{+\infty}_0 dV
  \left[ \ln \left({V\over 2b}\right) + \kappa_2\right] M^{(4)}_{ij}
   (T-V) + O(5)\ , \label{eq:4.32a} \\
 U_{ijk} (T) &=& M^{(3)}_{ijk} (T) + {2GM\over c^3} \int^{+\infty}_0 dV
  \left[ \ln \left({V\over 2b}\right) + \kappa_3\right] M^{(5)}_{ijk}
   (T-V) + O(5)\ , \label{eq:4.32b} \\
 V_{ij} (T) &=& S^{(2)}_{ij} (T) + {2GM\over c^3} \int^{+\infty}_0 dV
  \left[ \ln \left({V\over 2b}\right) + \kappa'_2\right] S^{(4)}_{ij}
   (T-V) + O(5)\ . \label{eq:4.32c}
\end{eqnarray}
\end{mathletters}
The other radiative moments $U_{ijkm},\cdots ,V_{ijkmn}$ in (\ref{eq:4.27})
and (\ref{eq:4.28}) are equal, with the required precision, to the
corresponding $M^{(4)}_{ijkm},\cdots ,S^{(5)}_{ijkmn}$. Three purely
numerical constants $\kappa_2$, $\kappa_3$ and $\kappa'_2$ appear in
(\ref{eq:4.32}), which are in factor of ``instantaneous" (non-hereditary)
contributions.  Note that there is a priori also a contribution
involving the interaction between the mass dipole $M_i$ and the mass
quadrupole $M_{jk}(T)$ in the octupole moment $U_{ijk}(T)$ of (4.32b).
This contribution, which is necessarily ``instantaneous" and of the type
$(\sigma_3 G/c^3) M_{\langle i} M^{(4)}_{jk\rangle} (T)$, where
$\sigma_3$ is some numerical constant, has been set to zero in
(\ref{eq:4.32b}) by requiring that the (harmonic) exterior coordinate
system is mass-centred, i.e.  $M_i=0$.  The computation of $\kappa_2$,
$\kappa_3$ and $\kappa'_2$ necessitates the implementation of the
algorithm for the construction of the external metric.  This was already
done in \cite{BD92} for the computation of $\kappa_2$.  The computation
done in the appendix~\ref{sec:apc} yields the values
\begin{equation}
  \kappa_2 = {11\over 12} \ , \qquad\quad
  \kappa_3 = {97\over 60} \ ,\qquad\quad
  \kappa'_2 = {7\over 6} \ . \label{eq:4.33}
\end{equation}
The constant $b$ entering the tail contributions in (\ref{eq:4.32}) is
a constant (with dimension of a time) which parametrizes the relation
between the radiative coordinate system $(T,R)$ in which the metric is
of the Bondi type and the harmonic
coordinate system $(t_{\rm can}, r_{\rm can})$ of \S~\ref{ssec:2.2}.  It
is such that
\begin{equation}
 T - {R\over c} = t_{\rm can} - {r_{\rm can}\over c} - {2GM\over c^3}
 \ln \left( {r_{\rm can}\over cb} \right) + O(5) \label{eq:4.34}
\end{equation}
(where terms of order $O(1/r^2_{\rm can})$ in the distance to the source
are neglected). A possibly convenient choice for the constant $b$ is
$b\sim 1/\omega_0$, where $\omega_0$ is a typical frequency at
which some detector at large distances from the source is operating
\cite{BSat94b}.

 The relations (\ref{eq:4.32}) are still not expressed in terms of the
source's parameters, and the last step obviously consists  in using the
relations (\ref{eq:4.7}) linking the ``canonical" moments $M_L$,
$S_L$ to the real source moments $I_L$, $J_L$. We can thus rewrite
(\ref{eq:4.32}) as
\begin{mathletters}
\label{eq:4.35}
\begin{eqnarray}
 U_{ij} (T) &=& I^{(2)}_{ij} (T) + {2GM\over c^3} \int^{+\infty}_0 dV
  \left[ \ln \left({V\over 2b}\right) + {11\over 12}\right] I^{(4)}_{ij}
   (T-V) + O(5)\ , \label{eq:4.35a} \\
 U_{ijk} (T) &=& I^{(3)}_{ijk} (T) + {2GM\over c^3} \int^{+\infty}_0 dV
  \left[ \ln \left({V\over 2b}\right) + {97\over 60}\right] I^{(5)}_{ijk}
   (T-V) + O(5)\ , \label{eq:4.35b} \\
 V_{ij} (T) &=& J^{(2)}_{ij} (T) + {2GM\over c^3} \int^{+\infty}_0 dV
  \left[ \ln \left({V\over 2b}\right) + {7\over 6}\right] J^{(4)}_{ij}
   (T-V) + O(4)\ , \label{eq:4.35c}
\end{eqnarray}
\end{mathletters}
(with relations limited to the first term in the right-hand-side for the
higher-order moments $U_{ijkm},\cdots$). One must insert these
relations, together with the explicit expressions (\ref{eq:4.21}) and
(\ref{eq:4.13}) of the source moments, into the waveform (\ref{eq:4.27})
and/or the energy-loss formula (\ref{eq:4.28}). (Note that the only tail
contribution in the energy-loss formula (\ref{eq:4.28}) comes from the
``mass-quadrupole" tail associated with the moment $U_{ij}$
in (\ref{eq:4.35a}).) This solves the problem
of the generation of gravitational waves by a general isolated system
at the 2-PN approximation.

\acknowledgments
 It is a pleasure to thank Kip S. Thorne and Clifford M. Will for
discussions raising the author's interest in this problem.

\appendix
\section{PN expansion of part of the external field}
\label{sec:apa}

 The ``$q$-part'' of the external metric is the second term in the
definition of the nonlinear canonical field (\ref{eq:2.21}). We shall
denote the first term in (\ref{eq:2.21}) by
\begin{equation}
 p^{\mu\nu}_{{\rm can}(n)} = {\rm FP}_{B=0} \Box^{-1}_{R} [r^B
     \Lambda^{\mu\nu}_{{\rm can}(n)}] \ . \label{eq:A1}
\end{equation}
The ``$q$-part'' of the metric is computed from the ``$p$-part''
(\ref{eq:A1}) as follows \cite{BD86}. We compute the divergence of
(\ref{eq:A1}),
namely $r^\mu_{{\rm can}(n)} = \partial_\nu p^{\mu\nu}_{{\rm can}(n)}$, and
obtain
\begin{equation}
 r^\mu_{{\rm can}(n)} = {\rm FP}_{B=0} \Box^{-1}_R [Br^{B-1} n_i
    \Lambda^{\mu i}_{{\rm can}(n)} ]\ ,  \label{eq:A2}
\end{equation}
where $n_i = x^i/r$ and where the factor $B$ comes from the derivation
of the analytic continuation factor $r^B$. This divergence is known
to be a retarded solution of the wave equation, and can thus be decomposed,
in a unique manner, as
\begin{mathletters}
\label{eq:A3}
 \begin{eqnarray}
 r^0_{{\rm can}(n)} &=& \sum_{\ell \geq 0} \partial_L \left[ {1\over r} A_L
    \left( t -{r\over c}\right) \right]\ , \label{eq:A3a} \\
 r^i_{{\rm can}(n)} &=& \sum_{\ell \geq 0} \partial_{iL} \left[ {1\over r} B_L
    \left( t -{r\over c}\right) \right]\  \nonumber \\
    &+& \sum_{\ell \geq 1} \left\{ \partial_{L-1} \left[ {1\over r} C_{iL-1}
    \left( t -{r\over c}\right) \right]
 + \varepsilon_{iab} \partial_{aL-1} \left[ {1\over r} D_{bL-1}
    \left( t -{r\over c}\right) \right] \right\}\ , \label{eq:A3b}
\end{eqnarray}
\end{mathletters}
where $A_L$, $B_L$, $C_L$ and $D_L$ are some STF tensorial functions of
the retarded time. Then the ``$q$-part'' of the external metric is
defined by its components as
\begin{mathletters}
\label{eq:A4}
\begin{eqnarray}
 q^{00}_{{\rm can}(n)} &=& -{c\over r} A^{(-1)} - c\partial_a \left( {1\over r}
  A^{(-1)}_a\right)+ c^2 \partial_a \left({1\over r} C_a^{(-2)}\right)\ ,
  \label{eq:A4a} \\
 q^{0i}_{{\rm can}(n)} &=& -{c\over r} C_i^{(-1)} - c\varepsilon_{iab}
  \partial_a \left( {1\over r} D^{(-1)}_b\right)- \sum_{\ell\geq 2}
  \partial_{L-1} \left({1\over r} A_{iL-1}\right)\ , \label{eq:A4b} \\
q^{ij}_{{\rm can}(n)} &=& -\delta_{ij} \left[ {1\over r} B + \partial_a
  \left( {1\over r} B_a\right)\right] \nonumber \\
 &+& \sum_{\ell\geq 2} \biggl\{ \partial_{L-2} \left( {1\over rc}
    A^{(1)}_{ijL-2} + {3\over rc^2} B^{(2)}_{ijL-2} - {1\over r}
   C_{ijL-2} \right) \nonumber \\
 &&\qquad +2\delta_{ij}\partial_L \left( {1\over r} B_L\right)
 - 6\partial_{L-1(i} \left( {1\over r} B_{j)L-1} \right)
    -2\partial_{aL-2} \left( \varepsilon_{ab(i} {1\over r}
  D_{j)bL-2} \right) \biggr\}\ , \label{eq:A4c}
\end{eqnarray}
\end{mathletters}
(where e.g. $A^{(-1)}(t) = \int^t_{-\infty} dt' A(t')$).

 The main task is to deal with the quadratic case $n=2$. We first
control the post-Newtonian expansion of the divergence
$r^\mu_{\rm can(2)}$ of (\ref{eq:A2}). The quadratic source
$\Lambda^{\mu\nu}_{\rm can(2)} = N^{\mu\nu}(h_{\rm can(1)})$ is computed
by inserting the linear metric (\ref{eq:2.17})-(\ref{eq:2.18}) into
(\ref{eq:1.5}). Its post-Newtonian expansion starts at $O(4,5,4)$, with
a next term at $O(6,7,6)$. Let us write
\begin{equation}
 \Lambda^{\mu\nu}_{\rm can(2)} = {1\over c^{4+\omega}} F^{\mu\nu}
   + {1\over c^{6+\omega}} G^{\mu\nu} + O(7,8,7)\ , \label{eq:A5}
\end{equation}
where $F^{\mu\nu}$ and $G^{\mu\nu}$ are the coefficients of the
leading-order and next-order terms in the post-Newtonian expansion, and where
we use
the notation $\omega =0$ when $\mu\nu =00$ or $ij$ and $\omega =1$ when
$\mu\nu =0i$. Then it is easy to show that the structures of
$F^{\mu\nu}$ and $G^{\mu\nu}$, as concerns their spatial dependence, are
\begin{mathletters}
\label{eq:A6}
\begin{eqnarray}
 F^{\mu\nu} &=&\sum_{p+q\geq 2} \partial_P \left({1\over r}\right)\partial_Q
  \left({1\over r}\right) \ , \label{eq:A6a}\\
 G^{\mu\nu} &=&\sum_{p+q\geq 1}\partial_P \left({1\over r}\right)
   \partial_Q \left({1\over r}\right) + \sum_{p+q\geq 2} \partial_P
   \left({1\over r}\right) \hat\partial_Q (r) \ ,
      \label{eq:A6b}
\end{eqnarray}
\end{mathletters}
where $P$ and $Q$ are multi-indices with $p$ and $q$ indices. Important
for our purpose is the fact that the number of spatial derivatives is
$p+q\geq 2$ in $F^{\mu\nu}$ and in the second term in $G^{\mu\nu}$, and
is $p+q\geq 1$ in the first term in $G^{\mu\nu}$. Knowing the expansion
(\ref{eq:A5}) we can write the corresponding expansion of the divergence
(\ref{eq:A2}). We have
\begin{eqnarray}
 r^\mu_{\rm can(2)} &=& {1\over c^{4+\omega}} {\rm FP}_{B=0} \Delta^{-1}
  [ Br^{B-1} n_i F^{\mu i} ] \nonumber \\
   &+& {1\over c^{6+\omega}} {\rm FP}_{B=0} \left\{ \Delta^{-1}
  [ Br^{B-1} n_i G^{\mu i} ] + \left( {\partial\over \partial t} \right)^2
  \Delta^{-2} [ Br^{B-1} n_i F^{\mu i} ] \right\} + O(8,7)\ ,
 \label{eq:A7}
\end{eqnarray}
where $\Delta^{-1}$ is the Poisson operator and $\Delta^{-2}=(\Delta^{-1})^2$,
and where $\omega =1$ when $\mu =0$ and $\omega =0$ when $\mu =i$. The
justification of the equation (\ref{eq:A7}) can be found in our previous
papers (see e.g. (3.25) in \cite{B93}). Using the structures (\ref{eq:A6})
of $F^{\mu\nu}$ and $G^{\mu\nu}$ into (\ref{eq:A7}) shows that all explicit
terms in (\ref{eq:A7}) are zero. Indeed, by multiplying by $r^{B-1}n_i$
the term $\partial_P (r^{-1}) \partial_Q (r^{-1})$ in (\ref{eq:A6a}) or
(\ref{eq:A6b}) and projecting on STF tensors, we get a series of terms
of the type $r^{B-\ell-2k-2}\hat n_L$ where $k$ is a positive or zero
integer and $\ell =p+q+1-2k$.  By applying $\Delta^{-1}$ and
$\Delta^{-2}$ we obtain $r^{B-\ell-2k}\hat n_L/D_{1,2} (B)$ where the
denominators are respectively $D_1(B)=(B-2\ell -2k)(B-2k+1)$ and
$D_2(B)=(B-2\ell -2k+2)(B-2\ell-2k)(B-2k+1)(B-2k+3)$.  When $p+q\geq 1$
(which implies $\ell +k\geq 1)$ neither $D_1(B)$ nor $D_2(B)$ vanish at
$B=0$, and when $p+q\geq 2$ (which implies $\ell +k\geq 2)$ $D_2(B)$
does not vanish either.  Furthermore, by multiplying by $r^{B-1}n_i$ the
second term $\partial_P(r^{-1})\hat\partial_Q (r)$ in (\ref{eq:A6b}),
projecting on STF tensors and applying $\Delta^{-1}$, we get
$r^{B-\ell-2k+2}\hat n_L/D_3(B)$ where the denominator is $D_3(B) =
(B-2\ell-2k+2)(B-2k+3)$ which does not vanish at $B=0$ when $p+q\geq 2$.
Since all denominators $D_1$, $D_2$, $D_3$ take non-zero values at $B=0$
we conclude that all terms will be zero at $B=0$ thanks to the explicit
factors $B$ present in (\ref{eq:A7}). Thus,
\begin{equation}
 r^\mu_{\rm can(2)} = O (8,7)\ . \label{eq:A8}
\end{equation}
The equation (\ref{eq:A3}) then shows that for $n=2$ the function $A_L$
is $O(8)$ while $B_L$, $C_L$ and $D_L$ are $O(7)$.  Hence we can write,
from (\ref{eq:A4}),
\begin{mathletters}
\label{eq:A9}
\begin{eqnarray}
 q^{00}_{\rm can(2)} &=& c^2
\partial_a \left( {1\over r} C^{(-2)}_a\right)+ O(7)\ , \label{eq:A9a}\\
 q^{0i}_{\rm can(2)} &=& -{c\over r} C^{(-1)}_i - c \varepsilon_{iab}
\partial_{a} \left({1\over r} D^{(-1)}_b\right)+ O(8)\ ,\label{eq:A9b} \\
 q^{ij}_{\rm can(2)} &=& O(7)\ . \label{eq:A9c}
\end{eqnarray}
\end{mathletters}
Thus it remains to control three terms involving antiderivatives of vectors
$C_i$ and $D_i$ and having low multipolarities $\ell =0,1$. We know that
the dependence on $c^{-1}$ of a term with multipolarity $\ell$ in
$q^{\mu\nu}_{\rm can(2)}$ is $O(5+\underline{\ell}_1+\underline{\ell}_2
-\ell)$ (see e.g.  (3.23) in \cite{B93}), where $\underline{\ell}_1$ and
$\underline{\ell}_2$ are the number of indices on the two moments ${\cal
P}_{\underline{L}_1}$ and ${\cal P}_{\underline L_2}$ composing the term
(notation of \S~\ref{sec:4.3}). Now one of  the two moments is necessarily
non-static since for stationary metrics the ``$q$-part'' of the metric
is zero (appendix~C in \cite{BD86}), thus $\underline{\ell}_1\geq 2$ say.
On the other hand, to form a vector $C_i$ or $D_i$ one needs
$\underline{\ell}_2 = \underline{\ell}_1 \pm 1$ thus $\underline{\ell}_1
+ \underline{\ell}_2 \geq 3$. This, together with the fact that $\ell \leq 1$,
shows that the remaining terms in (\ref{eq:A9}) are $O(7)$ at least.
Thus we have proved
\begin{equation}
 q^{\mu\nu}_{\rm can(2)} = O(7,7,7)\ . \label{eq:A10}
\end{equation}

We finally deal with the cubic case $n=3$. In this case we know that the
post-Newtonian expansion of the source $\Lambda^{\mu\nu}_{\rm can(3)}$
starts at $O(6,7,8)$.  [The fact that the spatial components $ij$ of the
cubic source are $O(8)$ instead of the expected $O(6)$ is not obvious
but has been proved in \cite{BD86} ---~see the proof
that $\tilde A=0$ on p.424 in \cite{BD86};  indeed a possible term
$O(6)$ would be made of three mass-type multipoles $M_L$.] Thus the
divergence (\ref{eq:A2}) with $n=3$ is at least $r^\mu_{\rm can(3)} =
O(7,8)$, from which we deduce $q^{\mu\nu}_{\rm can(3)} = O(6,7,8)$.  On
the other hand the dependence in $c^{-1}$ of a term in $q^{\mu\nu}_{\rm
can(3)}$ is $O(8+\Sigma \underline\ell_i -\ell)$ (see (3.23) in
\cite{B93}) and we know that $\Sigma\underline\ell_i \geq \ell -s$ by
the law of addition of angular momenta, from which we deduce also
$q^{\mu\nu}_{\rm can(3)} = O(8-s) = O(8,7,6)$.  These two results imply
\begin{equation}
  q^{\mu\nu}_{\rm can(3)} = O(8,7,8)\ . \label{eq:A11}
\end{equation}
Equations (\ref{eq:A10}) and (\ref{eq:A11}) are the ones which are used
in the text.

\section{The conserved 2-PN total mass}
\label{sec:apb}

 We first obtain the expression of the total conserved mass at the 2-PN
approximation. The equation of continuity at this level of approximation reads
as
\begin{eqnarray}
&&\partial_t \left[ \sigma \left( 1 - 4P/c^4\right) \right] +
\partial_j \left[ \sigma_j \left( 1 - 4P/c^4 \right) \right] \nonumber\\
&& \qquad = {1\over c^2} (\partial_t \sigma_{jj} -\sigma\partial_t V) -
 {4\over c^4} (\sigma U_j \partial_j U + \sigma_{jk} \partial_j U_k)
 + O(5)\ , \label{eq:B1}
\end{eqnarray}
where our notation can be found in (\ref{eq:2.1})-(\ref{eq:2.4}) and
(\ref{eq:4.14})-(\ref{eq:4.15}). By integrating this equation  over the
three dimensional space we obtain
\begin{eqnarray}
 &&{d\over dt} \left[ \int d^3{\bf y} \left\{ \sigma -{1\over c^2}
  \sigma_{jj} - {4\over c^4} \sigma P\right\} \right] \nonumber \\
 &&\qquad = \int d^3{\bf y} \left\{ -{1\over c^2}
  \sigma \partial_t V - {4\over c^4} (\sigma U_j\partial_j U + \sigma_{jk}
 \partial_j U_k )\right\} + O(5)\ .  \label{eq:B2}
\end{eqnarray}
 Several transformations of both sides of this equation yield the
equation of conservation of mass at the 2-PN approximation, namely
\begin{eqnarray}
 &&{d\over dt} \biggl[ \int d^3{\bf y} \biggl\{ \sigma +{1\over c^2} \left(
 - \sigma_{jj} + {1\over 2} \sigma V \right) \nonumber \\
 &&\qquad +{1\over c^4} \left( \sigma U^2 +2\sigma_i U_i -
 4 \sigma_{jj} U - {1\over 4} \partial_t\sigma \partial_t X \right)
  \biggr\}\biggr] = O(5)\ .  \label{eq:B3}
\end{eqnarray}

We now show that the expression (\ref{eq:4.21}) we obtained in the text
for the general mass-type source moment $I_L$ reduces when $\ell =0$ to
the conserved mass appearing in the square brackets of (\ref{eq:B3}).
When $\ell =0$ the expression (\ref{eq:4.21}) becomes
\begin{eqnarray}
 I =&&{\rm FP}_{B=0} \int d^3{\bf y}|{\bf y}|^B \biggl\{ \sigma +{4\over c^4}
 (\sigma_{ii}U -\sigma P) + {1\over 6c^2} |{\bf y}|^2\partial^2_t \sigma
 \nonumber\\
 &&-{4\over 3c^2} y_i \partial_t \left[ \left( 1+{4U\over c^2}\right) \sigma_i
  + {1\over \pi Gc^2}\left( \partial_k U[\partial_i U_k -\partial_k U_i]
  + {3\over 4} \partial_tU \partial_iU\right)\right]\nonumber \\
 &&+{1\over 120c^4} |{\bf y}|^4\partial^4_t \sigma - {2\over 15c^4}
  |{\bf y}|^2y_i\partial^3_t \sigma_i
 +{1\over 5c^4} \hat y_{ij}\partial^2_t \left[ \sigma_{ij} + {1\over 4\pi G}
   \partial_i U\partial_j U\right] \nonumber \\
 &&+{1\over \pi Gc^4} \biggl[ -P_{ij}\partial^2_{ij} U-2U_i\partial_t
  \partial_iU +2\partial_iU_j\partial_jU_i
    - {3\over 2} (\partial_t U)^2 - U\partial^2_t U\biggr]
 \biggr\} + O(5)\ . \label{eq:B5*}
\end{eqnarray}
We shall not write down the rather long calculation but only indicate
its main steps. One must use the equation of continuity (\ref{eq:B1})
at the 1-PN approximation, i.e.
\begin{mathletters}
\label{eq:B4}
\begin{equation}
 \partial_t \sigma + \partial_j\sigma_j = {1\over c^2} (\partial_t \sigma_{jj}
 - \sigma \partial_t U) + O(4)\ , \label{eq:B4a}
\end{equation}
together with the corresponding equation of motion
\begin{eqnarray}
&&\partial_t \left[\sigma_i \left( 1+4U/c^2\right)\right] +
 \partial_j \left[ \sigma_{ij} \left( 1+4U/c^2\right)\right]\nonumber\\
&&\qquad = \sigma\partial_i V+{4\over c^2} [ \sigma\partial_t U_i
 + \sigma_j (\partial_jU_i -\partial_iU_j) ] + O(4)\ . \label{eq:B4b}
\end{eqnarray}
\end{mathletters}
Thanks to the lemma (\ref{eq:4.2}) we know that the non-compact supported
terms in (\ref{eq:B5*}) can be freely integrated by parts as if the
analytic continuation factor $|{\bf y}|^B$ and the finite part prescription
were absent, and as if all surface terms were zero. Let us quote here a
list of identities, valid up to the required precision and up to the
addition of a total divergence, which are used in the
reduction of (\ref{eq:B5*}).
\begin{mathletters}
\label{eq:B5}
\begin{eqnarray}
 &&{4\over c^4} (\sigma_{ii} U-\sigma P) = {1\over c^4} \sigma U^2 \ ;
 \label{eq:B5a}\\
 &&-{1\over \pi Gc^4} P_{ij} \partial^2_{ij} U =
  - {1\over \pi Gc^4} U\partial_t^2 U\ ; \label{eq:B5b}\\
 && {1\over 120c^4} |{\bf y}|^4 \partial^4_t\sigma =
   {1\over 30c^4} |{\bf y}|^2 y_i\partial^3_t\sigma_i \ ; \label{eq:B5c}\\
 && {1\over 5c^4} \hat y_{ij}\partial^2_t \left[ \sigma_{ij}
   + {1\over 4\pi G} \partial_i U\partial_j U\right] =
   {1\over 10c^4} |{\bf y}|^2 y_i\partial^3_t \sigma_i
   - {1\over 6c^4} |{\bf y}|^2 \partial^2_t \left[ \sigma_{ii}
   - {1\over 8\pi G} \partial_i U\partial_i U\right] \ ; \label{B5d} \\
 && - {4\over 3\pi Gc^4} y_i\partial_t [\partial_k U(\partial_i U_k
   - \partial_kU_i)] = {16\over 3c^4} y_i\partial_t \left[ U\sigma_i
   - {1\over 4\pi G} U \partial_i \partial_t U\right]\ . \label{eq:B5e}
\end{eqnarray}
\end{mathletters}
Using these and other identities, we arrive finally at a manifestly
compact-supported expression (on which we can remove the analytic
continuation factors) which reads as
\begin{eqnarray}
 I =&& \int d^3{\bf y} \biggr\{ \sigma + {1\over c^2} \left( -\sigma_{jj}
 + {1\over 2} \sigma V \right) \nonumber\\
 && \qquad + {1\over c^4} \left( \sigma U^2 +2\sigma_iU_i - 4\sigma_{jj} U
  -{1\over 4} \partial_t \sigma \partial_t X\right) \biggr\} +O(5)\ ,
  \label{eq:B6}
\end{eqnarray}
in perfect agreement with (\ref{eq:B3}).

\section{Computation of three constants}
\label{sec:apc}

To compute the three constants $\kappa_2$, $\kappa_3$ and $\kappa'_2$
appearing in (\ref{eq:4.32}) one needs to implement the construction
of the external metric for the interacting multipoles $M\times M_{ij}$
(case of $\kappa_2)$, $M\times M_{ijk}$ (case of $\kappa_3$) and
$M\times S_{ij}$ $(\kappa'_2)$. The more general cases of interacting
multipoles $M\times M_L(t)$ and $M\times S_L (t)$ are in fact not more
difficult to handle and probably will be useful in future work, so
we shall compute $\kappa_\ell$ and $\kappa'_\ell$
for any $\ell \geq 2$.

 In the appendix~B of \cite{BD92}, where $\kappa_2 =11/12$ was obtained,
we computed the complete $M\times M_{ij}$ metric valid all over $D_e$.
Here we shall only compute the terms $1/r$ (and $\ln r/r$) in the
$M\times M_L$ and $M\times S_L$ metrics at large distances from the
source, since $\kappa_\ell$ and $\kappa'_\ell$ are contained in these
terms.  The quadratic source (\ref{eq:1.5}) of Einstein's equations,
computed with the linearized metric (\ref{eq:2.17})-(\ref{eq:2.18}) and
in which we retain only the products of multipoles $M\times M_L (t)$ and
$M\times S_L(t)$, is made of a series of terms of the type $\partial_P
(r^{-1}) \hat\partial_Q (r^{-1} F(t-r/c))$, where the function $F(t)$ is
some time-derivative of a moment $M_L(t)$ or $S_L(t)$ and where the
number of space-derivatives acting on $r^{-1}$ is at most two, i.e.
$p=0,1$ or $2$ (and where $q$ is arbitrary).  Thus we need only to
compute the leading term when $r\to +\infty$ of the (finite part of
the) retarded integral of $\partial_P (r^{-1}) \hat \partial_Q (r^{-1}
F(t-r/c))$ when $p=0,1$ or 2.  When $p=0$ we know from our previous
papers that this retarded integral involves a tail.  A computation using
(2.26) in \cite{BD92} and (4.24) in \cite{BD88} leads to
\begin{eqnarray}
 {\rm FP}_{B=0} \Box^{-1}_R \left[ r^{B-1}\hat\partial_Q \left(
  r^{-1} F \left( t-{r\over c} \right)
 \right)\right] =&&{(-)^q\hat n_Q\over 2rc^q} \int^\infty_0 dx F^{(q)}
 \left( t-{r\over c} -{x\over c}\right) \nonumber \\
  &&\times\left[ \ln \left( {x\over 2r}\right)
 + \sum^q_{k=1} {1\over k} \right] + O\left( {\ln r\over r^2}\right)\ .
 \label{eq:C1}
\end{eqnarray}
(Note that the sum $\Sigma^q_{k=1} (1/k)$ is multiplied by a factor 1
in the present formula (\ref{eq:C1}) and by a factor 2 in the formula
(2.26) in \cite{BD92}.) When $p=1$ or 2 there are no tails but the
calculation is in fact somewhat more complicated.  One must use (4.26)
in \cite{BD88} to get the polar part at $B=0$ of some integrals.  The
results are
\begin{eqnarray}
 {\rm FP}_{B=0} \Box^{-1}_R&&\left[ r^B \partial_i (r^{-1}) \hat\partial_Q
\left( r^{-1} F \left( t-{r\over c}\right) \right)\right]  \nonumber\\
 &&= {(-)^q\over 2(q+1)} (n_i\hat n_Q -\delta_{i\langle a_q} n_{Q-1\rangle})
 {F^{(q)} \left( t-{r\over c}\right)\over rc^q} + O\left( {1\over r^2}\right)
  \ , \label{eq:C2}\\   && \nonumber \\
 {\rm FP}_{B=0} \Box^{-1}_R&&\left[ r^B\partial_{ij} (r^{-1}) \hat\partial_Q
  \left( r^{-1} F \left( t-{r\over c}\right) \right)\right]
  = {(-)^{q+1}\over 2(q+1)(q+2)} \biggl\{ (n_{ij}+\delta_{ij})\hat n_Q
  \nonumber \\
 && -2 [\delta_{i\langle a_q} n_{Q-1\rangle}  n_j
    +\delta_{j\langle a_q} n_{Q-1\rangle}n_i]
   + 2\delta_{i\langle a_q}\delta_{\underline{j}a_{q-1}} n_{Q-2\rangle}
  \biggr\} {F^{(q+1)} \left( t-{r\over c}\right)\over rc^{q+1}}
    + O\left( {1\over r^2}\right) \ . \label{eq:C3}
\end{eqnarray}
(We denote $Q=a_1\cdots a_q$, and in the last term of (\ref{eq:C3}),
$\underline{j}$  means that $j$ has to be excluded from the STF operation
$\langle\ \rangle$.) It is then straightforward to obtain
the needed nonlinear sources with the help of (\ref{eq:1.5}) and to
apply the formulas (\ref{eq:C1})-(\ref{eq:C3}) on each terms of these
sources.  The ``$p$-part'' of the metric obtained in this way (see
(\ref{eq:A1})) is found to be divergenceless up to order $O(\ln r/r^2)$
and thus, as a short reasoning shows, the $``q$-part'' of the metric
(see (\ref{eq:A4})) vanishes at this order.  In the case of the
interacting multipoles $M\times M_{L}$, we find the metric
\begin{mathletters}
\label{eq:C4}
\begin{eqnarray}
 h^{00}_{\rm can(2)} &=&{8\over \ell !(\ell-1)} {n_LMM^{(\ell+1)}_L\over r}
 - {8\over \ell !} {n_LM\over r} \int^\infty_0 dx M^{(\ell+2)}_L
 \left[ \ln \left( {x\over 2r}\right) + \sum^\ell_{k=1} {1\over k} \right]
 + O \left( {\ln r\over r^2}\right)\ , \label{eq:C4a}\\
 h^{0i}_{\rm can(2)} &=&{-2\over\ell !(\ell+1)}{n_{iL}MM^{(\ell+1)}_L\over r}
 + {2\over \ell !} {\ell^2+3\ell+4\over \ell(\ell+1)(\ell-1)}
  {n_{L-1}MM^{(\ell+1)}_{iL-1}\over r} \nonumber\\
 &&- {8\over \ell !} {n_{L-1}M\over r} \int^\infty_0 dx M^{(\ell+2)}_{iL-1}
 \left[ \ln \left( {x\over 2r}\right) + \sum^{\ell-1}_{k=1} {1\over k} \right]
 + O \left( {\ln r\over r^2}\right)\ , \label{eq:C4b}\\
 h^{ij}_{\rm can(2)}&=&-{4\over \ell !(\ell+2)} {n_{ijL}MM^{(\ell+1)}_L\over r}
 + {4\over \ell !} {5\ell^2+10\ell+8\over \ell(\ell+1)(\ell+2)}
  {n_{L-1(i}MM^{(\ell+1)}_{j)L-1}\over r} \nonumber\\
 && - {4\over \ell !} {2\ell^2+5\ell+4\over \ell(\ell+1)(\ell+2)}
  {\delta_{ij}n_LMM^{(\ell+1)}_L\over r}
  - {8\over \ell !} {2\ell^2+5\ell+4\over \ell(\ell+1)(\ell+2)}
  {n_{L-2}MM^{(\ell+1)}_{ijL-2}\over r} \nonumber \\
 &&- {8\over \ell !} {n_{L-2}M\over r} \int^\infty_0 dx M^{(\ell+2)}_{ijL-2}
\left[ \ln \left( {x\over 2r}\right) + \sum^{\ell-2}_{k=1} {1\over k}
\right]
 + O \left( {\ln r\over r^2}\right)\ , \label{eq:C4c}
\end{eqnarray}
\end{mathletters}
(with $c=1)$. When $\ell =2$ this metric reduces to (B4) in the
appendix~B of \cite{BD92}. In the case of the interacting multipoles
$M\times S_L$, we find
\begin{mathletters}
\label{eq:C5}
\begin{eqnarray}
 h^{00}_{\rm can(2)} &=& O\left( {\ln r\over r^2}\right)\ , \label{eq:C5a}\\
 h^{0i}_{\rm can(2)} &=& - {8(\ell+2)\over (\ell+1)!(\ell+1)}
 {n_{aL-1}\varepsilon_{iab} MS^{(\ell+1)}_{bL-1}\over r}\nonumber\\
  &&+{8\ell\over (\ell+1)!} {n_{aL-1}\varepsilon_{iab} M\over r}
   \int^\infty_0 dx S^{(\ell+2)}_{bL-1}
 \left[\ln \left( {x\over 2r}\right) + \sum^\ell_{k=1} {1\over k}\right]
  + O\left( {\ln r\over r^2}\right)\ , \label{eq:C5b}\\
 h^{ij}_{\rm can(2)} &=& - {16\ell\over (\ell+1)!(\ell+1)}
 {n_{aL-1(i}\varepsilon_{j)ab} MS^{(\ell+1)}_{bL-1}\over r}\nonumber\\
  &&+ {16(\ell-1)\over (\ell+1)!(\ell+1)}
 {n_{aL-2}\varepsilon_{ab(i} MS^{(\ell+1)}_{j)bL-2}\over r}\nonumber\\
 &&+ {16\ell\over (\ell+1)!} {n_{aL-2}\varepsilon_{ab(i} M\over r}
   \int^\infty_0 dx S^{(\ell+2)}_{j)bL-2}
 \left[\ln \left( {x\over 2r}\right) + \sum^{\ell-1}_{k=1} {1\over k}\right]
  + O\left( {\ln r\over r^2}\right)\ . \label{eq:C5c}
\end{eqnarray}
\end{mathletters}
In (\ref{eq:C4}) and (\ref{eq:C5}) the moments are evaluated at $t-r/c$
in the instantaneous terms, and at $t-r/c-x/c$ in the tail terms.
{}From the metrics (\ref{eq:C4}) and (\ref{eq:C5}) we immediately deduce the
values of $\kappa_\ell$ and $\kappa'_\ell$ entering the tail terms in the
radiative moments. These are
\begin{eqnarray}
 \kappa_\ell &=& {2\ell^2 +5\ell+4\over \ell(\ell+1)(\ell+2)} +
 \sum^{\ell-2}_{k=1} {1\over k} \ , \label{eq:C6}\\
 \kappa'_\ell &=& {\ell-1\over \ell(\ell+1)} +
 \sum^{\ell-1}_{k=1} {1\over k} \ . \label{eq:C7}
\end{eqnarray}
We thus find the values quoted in (\ref{eq:4.33}). Note that the constants
$\kappa_\ell$ and $\kappa'_\ell$ depend on the coordinate system which is
used, namely the harmonic coordinate system. For instance, the constants
would be $\kappa_\ell +{1\over 2}$ and $\kappa'_\ell +{1\over 2}$ in a
(perturbed) Schwarzschild coordinate system. Note also that
\begin{mathletters}
 \label{eq:C8}
\begin{eqnarray}
 \lim_{\ell\to+\infty} [\kappa_\ell - \ln \ell] &=& C \label{eq:C8a}\\
 \lim_{\ell\to+\infty} [\kappa'_\ell - \ln \ell] &=& C\ , \label{eq:C8b}
\end{eqnarray}
\end{mathletters}
where $C=0.577$\dots\ is Euler's constant. This can be of interest since
the combinations $\kappa_\ell -\ln \ell -C$ and $\kappa'_\ell -\ln \ell
-C$ arise in the phase of the Fourier transform of the waveform (see
(3.5) in \cite{BS93} for the mass-quadrupole case $\ell =2$).

\end{document}